\documentclass[]{jfm}

\usepackage{graphicx}
\usepackage{newtxtext}
\usepackage{newtxmath}
\usepackage{natbib}
\usepackage{hyperref}
\hypersetup{
    colorlinks = true,
    urlcolor   = blue,
    citecolor  = black,
}

\newcommand{\RomanNumeralCaps}[1]

\usepackage{xcolor,soul}
\usepackage{soulutf8}

\soulregister\citep7
\soulregister\cite7
\soulregister\ref7
\soulregister\eqref7

\title{Finite-time mixing and reaction in packed bed reactors}

\author{Dario Maggiolo\aff{1}
  \corresp{\email{maggiolo@chalmers.se}},
  Lucien Magson \aff{2}
  Gunther Munz \aff{3}
  Diego Sampedro \aff{2}
 \and Angela Sasic Kalagasidis \aff{4}}

\affiliation{
\aff{1}Department of Mechanics and Maritime Sciences, Chalmers University of Technology, 412 96, Göteborg, Sweden
\aff{2}Departamento de Química, Instituto de Investigación Química de la Universidad de La Rioja, Universidad de la Rioja,
	C/Madre de Dios 53, 26006, Logroño, La Rioja, Spain
\aff{3}Heating and Cooling Technologies, Fraunhofer Institute for Solar Energy Systems (ISE), Heidenhofstrase 2, 79110, Freiburg, Germany
\aff{4}Department of Architecture and Civil Engineering, Chalmers University of Technology, 412 96, Göteborg, Sweden}

\begin{document}
\maketitle

\begin{abstract}
We investigate mixing dynamics in porous media at finite times, using pore-scale lattice–Boltzmann simulations combined with Lagrangian particle tracking. We compute fluid deformation in randomly packed beds based on the moving Protean frame approach introduced by~\cite{lester2018fluid}. From the extracted Lagrangian kinematics, we construct a mixing model based on lamellar aggregation that well predicts the Eulerian scalar fields obtained from simulations.
Our results reveal an early time mixing regime dominated by shear-driven fluid deformation, where solute mixing arises from the random overlap of diffusive concentration elements. In this regime, mixing proceeds slowly and follows a temporal decay of concentration variance, $\sigma_c^2 \propto \Pen^{-\alpha/(2\alpha+1)} t^{-1/2}$, where $\Pen$ is the Péclet number and $\alpha$ the exponent characterizing shear deformation. This dynamic arises when the Péclet number is small relative to the ratio between the exponential-mixing and shear-deformation timescales.
This analysis also demonstrates that shear-induced mixing govern the homogenization of early-stage reactions at the fluid–solid interface in finite-size random packed beds, typically operating at moderate Péclet numbers $\Pen =O(10^2)$. 
\end{abstract}

\begin{keywords}
Authors should not enter keywords on the manuscript, as these must be chosen by the author during the online submission process and will then be added during the typesetting process (see \href{https://www.cambridge.org/core/journals/journal-of-fluid-mechanics/information/list-of-keywords}{Keyword PDF} for the full list).  Other classifications will be added at the same time.
\end{keywords}

{\bf MSC Codes }  {\it(Optional)} Please enter your MSC Codes here

\section{Introduction}
\label{sec:intro}

Although fully developed mixing in porous media has been widely studied, the dynamics of mixing at early, finite times remains less well understood.

In porous systems of limited size or characterized by finite-time processes such as reactions, finite-time mixing plays a decisive role in controlling  the homogenization of an injected solution. Consider, for instance, small compact porous reactors: in such systems, while molecular diffusion is too slow to support mixing, pore-scale advection contributes to distributing the solute effectively within the limits of the reactor's longitudinal size $L$. Nevertheless, the residence time $L/U$ has to be long enough to allow complete molecular conversion~\citep{maggiolo2020solute}, implying that the device must operate at a limited mean flow velocity $U$ and Péclet number. As a result, advection-driven mixing is weak and constrained by the limited physical reactor space $L$.

Limited-size porous systems are also encountered in a range of other applications, such as membranes and electrodes~\citep{farzaneh2021pore}. Furthermore, in highly reactive media, mixing may be limited at very short times, because species transport is subsequently inhibited by reaction. Under such conditions, fully developed mixing may not be achieved and the mixing dynamics are fully realized at finite times.

Whether because of a fast reaction time or a physically short medium, the time available for solute mixing may be shorter than the characteristic mixing time $t_m$ that marks the onset of asymptotic behavior. This mixing time scales proportionally to the characteristic pore length $d$ and inversely to the mean flow velocity, $t_m \propto d/U$. More precisely, $t_m$ reflects the specific kinematic mechanisms of mixing dictated by the pore-scale flow and geometry.

At finite or large Péclet numbers, solutes transported through porous media often organize into thin, high-concentration filaments separated by low-concentration voids~\citep{dentz2015mixing}. Mixing refers to the transition of a scalar field--such as concentration--from an initially heterogeneous, segregated state to a final, spatially homogeneous distribution~\citep{villermaux2019mixing}. In porous media, this process relies on the ability of the flow to eliminate small-scale concentration contrasts through kinematic deformations generated by the pore-scale geometry at very early times. In other words, mixing is completely mediated by the pore-scale architecture.

Unlike two-dimensional flows, three-dimensional porous structures inherently generate chaotic flow~\citep{lester2013chaotic,lester2016chaotic}, where random sequences of stretching and folding events produce long-term exponential stretching of concentration fields~\citep{heyman2020stretching}. The average long-term growth rate of infinitesimal fluid element lengths is quantified by the Lyapunov exponent, $\lambda$, which determines the onset of chaotic mixing at $t_m \sim d/(2 \lambda U) \ln (\Pen)$. \cite{turuban2019chaotic} further demonstrated that chaotic mixing in granular porous media with simple artificial lattice structures--although weaker than in non-granular open pore networks--depends strongly on the orientation of the mean flow relative to the lattice geometry. For the specific case of random stretching orientations, the Lyapunov exponent has been theoretically estimated as $\lambda \approx 0.11~$\citep{lester2013chaotic,lester2016chaotic}. Experimental studies in random packed beds report slightly higher values: $\lambda = 0.20$~\citep{Kree2017scalar}, $\lambda = 0.21~$\citep{heyman2020stretching}, and $\lambda = 0.17$~\citep{heyman2021scalar}.

In the presence of fast reaction times, or if the porous bed physical length is limited to only a few grain diameters, there is insufficient time for fully developed mixing, since the residence time $1/k_r$ or $L/U$, respectively, becomes comparable to the mixing time $d/(\lambda U)$. The Lyapunov exponent is an infinite-time measure and does not fully capture the three-dimensional deformation. At such short times, shear-induced deformation and chaotic mixing coexist.
\cite{lester2018fluid} showed that at finite times, shear-induced deformation may dominate. \cite{heyman2021scalar} documented the emergence of chaotic mixing signatures only after an initial transient phase, during which concentration differences persist. This transient phase is consistent with the fact that chaotic mixing requires time to fully develop (at least the time needed to reach the Batchelor scale corresponding to the exponential mode of mixing). \cite{aquino2023fluid} showed how shear deformation governs the long-time distribution of species in porous media and the statistics of travel times to solid reactive surfaces, effectively determining the reaction efficiency of the porous medium. However, the transient regime leading to this asymptotic limit remains insufficiently explored.

Lamellar theories have been developed to capture these dynamics. Initially proposed for turbulent flows~\citep{duplat2008mixing} and later extended to random porous flows~\citep{dentz2023mixing}, these models describe mixing as the formation of elongated lamellae (or sheets) of high concentration and their subsequent random aggregation. This framework highlights an early stretching–compression regime, during which advection (i) redistributes regions of high concentration randomly in space, bringing previously segregated pockets into proximity, and (ii) amplifies transverse concentration gradients within lamellae, thereby accelerating diffusion-driven dilution. Both processes act as powerful mixing mechanisms. After $t_m$, the overlapping lamellae can be modeled as undergoing random aggregation, either fully or partially correlated depending on the pore-scale dynamics~\citep{duplat2010entanglement,heyman2024mixing}.

The lamellar model thus provides a useful framework for studying finite-time mixing, as it also captures the early-time fluid-induced deformation kinematics.
This kinematics depend on several factors, most notably (i) the spatial dimensionality of the stirring protocol and (ii) the distribution and randomness of the flow. \cite{villermaux2019mixing} reviews various stirring protocols, including the two-dimensional shear-induced deformation. For this case, the predictive accuracy of lamellar theory has been experimentally validated~\citep{souzy2018mixing}, and the mixing time has been shown to exhibit a weak algebraic dependence on the Péclet number, scaling as $t_m \sim (d/U) \Pen^{1/3}$. A similar scaling applies to shear-induced mixing by a vortex~\citep{villermaux2019mixing}.

In the present work, we quantitatively investigate the finite-time behavior of mixing in monodisperse packed beds using computational analyses within both Eulerian and Lagrangian frameworks. \\
The research questions we address are: (i) which mixing mode is most relevant at finite times in packed-bed porous media; (ii) how this mixing mode affects the pore-scale concentration distribution at the fluid–solid interface; and (iii) whether the homogeneity of surface reactions in finite-size packed-bed reactors can be modeled and predicted. From a fundamental perspective, addressing these questions is essential to advancing the understanding of early-time mixing in porous media. From a technological perspective, new models are required to capture the early-time redistribution of species at fluid–solid interfaces within packed beds, thereby improving the design and operational control of small-scale porous reactors.

Two numerical approaches are commonly proposed to quantify pore-scale mixing and to extract the corresponding macroscopic reaction rates. Eulerian models track the temporal evolution of concentration fields~\citep{gurung2020mixing}, whereas Lagrangian approaches, such as random reactive walks~\citep{valocchi2019mixing,sole2021random}, follow individual tracer trajectories. Both approaches face computational challenges: Eulerian models require high-resolution pore-scale discretization, while Lagrangian models are constrained by the number of tracers that can be resolved.
Other modeling frameworks, including fractal derivative diffusion models~\citep{liang2023characterization} and effective reactive transport equations~\citep{dentz2011effective}, can be used to account for heterogeneity, anomalous diffusion, and small-scale mixing. However, these models rely on microstructural parameters describing the pore geometry, which are often difficult to measure a priori, making direct pore-scale simulations indispensable.

To solve the Eulerian flow and concentration fields at the pore-scale we use the lattice Boltzmann methodology (LBM), which is well suited for solving laminar flows in complex geometry, since it allows an easy local representation of the pore surfaces projected onto a regular spatial grid. We consider as a test-case the geometry of a packed bed contained in tubular channels of finite length. We then apply a Lagrangian Particle Tracking (LPT) algorithm to calculate the Lagrangian trajectories of massless particles (tracers) advected by the flow. Along these trajectories, we compute the deformation tensor, based on the moving Protean framework of~\cite{lester2018fluid}, which ultimately allows us to determine the dominant modes of fluid-induced deformation at finite times. We then employ a lamellar model constructed over the computed deformations to predict the spatial reorganization and mixing of  regions of varying concentration. We assess the validity of the lamellar model by comparing its predictions with the solution of the Eulerian concentration field.

The paper is organized as follows: Section~\ref{sec:nummet} describes the numerical methodology, Sections~\ref{sec:top} and \ref{sec:def} investigate the characteristics of pore-scale velocities and deformations, Sections~\ref{sec:msd}, \ref{sec:mix} analyze the evolution of the concentration fields, whereas Sections~\ref{sec:sp} and \ref{sec:agg} interpret the mixing dynamics using a lamellar aggregation model, which we will then extend to reactive porous systems in Section~\ref{sec:rea}. Section~\ref{sec:con} concludes with a summary of the main findings.

\section{Numerical Methodology~\label{sec:nummet}}

\subsection{Geometry construction}

 \begin{figure}
  \centerline{\includegraphics[width=0.9\linewidth]{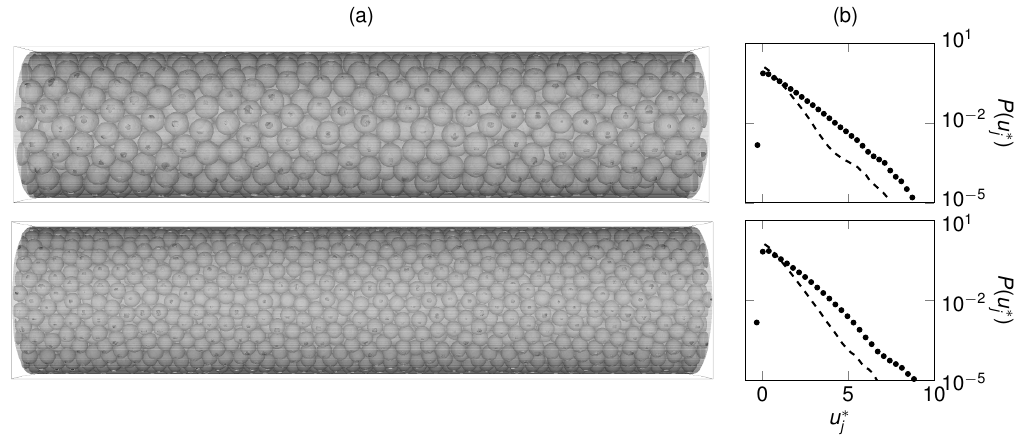}}
  \caption{(a) Two tubular reactors of radius $R$ containing the reactive porous media composed of catalyst spherical particles of size $d= R/3.1$ (upper panel) and $d=R/4.7$ (lower panel). The two systems are used as geometrical input  for the lattice-Boltzmann  simulations. Right panels (b): probability distribution functions of normalised flow velocities ${u_j^*(\boldsymbol{x})}=u_j(\boldsymbol{x})/U$ (with $U$ the mean streamwise flow velocity) along the streamwise $u_x$ (black marks) and transverse $u_t=(u_y^2+u_z^2)^{1/2}$ (dashed line) directions; particles of size $d= R/3.1$ (upper panel) and $d=R/4.7$ (lower panel). }
\label{fig1}
\end{figure}

To study finite-time mixing, we take a small fixed packed bed as an example: quasi-spherical catalytic particles are close-packed into a tubular reactor (figure~\ref{fig1}). A molecular solution is continuously injected, while the catalyst particle surfaces trigger chemical reactions that progressively convert the solute. A notable example is Molecular Solar Thermal Energy Storage Systems (MOST), where light-activated molecules store solar energy in the form of chemical bonds ~\citep{wang2019macroscopic,magson2024sust}. The solute concentration is typically kept low to prevent molecular inhibition, while the porous geometry of the packed bed provides a high specific surface area for reactions. The choice of a small reactor size offers advantages such as (i) efficient energy extraction, (ii) reduced catalyst use and maintenance, and (iii) compact integration into small containers.
These systems are highly sensitive to mixing evolution at the pore-scale. Indeed, assuming fully-developed mixing and a rapid development of well-mixed conditions can lead to significant errors in predicting reactor performance~\citep{sole2022closer}.

We construct the reactor geometry using the open-source computer graphics software tool Blender. The rigid-body dynamics of spherical particles falling into a cylindrical container are simulated to mimic the microstructure of catalyst reactors composed of spherical particles. We consider two geometrical configurations, with reactor radius to particle diameter ratios of $R/d=3.1$ and $4.7$. The reactor lengths are scaled accordingly, with $L=28d$ and $42d$. 

After particle deposition, we extract the particle center positions and reconstruct binarized three-dimensional image matrices of the resulting geometries. These binarized images contain $7.5 \ 10^7$ and $13.1 \ 10^8$ voxels, with voxel-to-particle size ratios of 
$\Delta x/d \approx 1/40$ and $\Delta x/d \approx 1/32$ for the $R/d=3.1$ and $4.7$ configurations, respectively. This resolution ensures that the voxel size, later used for the numerical lattice-Boltzmann computation, remains smaller than the smallest characteristic transport scale relevant to the targeted flow velocity (Batchelor length scale under linear shear $s_b/d \sim{\Pen}^{-1/3} \approx 1/7$, under exponential stretching in packed beds $s_b/d \sim (0.1\Pen)^{-1/2} \approx 1/5$, see also Section~\ref{sec:agg}). The two configurations are shown in figure~\ref{fig1}, panel (a).

\subsection{Pore-scale simulations}

The lattice-Boltzmann methodology (LBM) is applied to solve the momentum equation and solute reaction at the pore-scale. LBM is particularly well suited for this task due to its ease of algorithmic parallelization, enabling efficient numerical solvers on high-performance computing infrastructures, including flows in complex geometries. We use the open-source code lbdm~\citep{lbdm}, which has previously been applied to model solute transport and reactions in porous media. Validation of this algorithm for porous media applications is provided e.g. in \cite{maggiolo2023transition}.

The momentum equation is first solved within the porous geometry by computing the streaming and collision of a lattice population driven by a body force that mimics a pressure gradient. Two buffer zones of length $R$ are added at the inlet and outlet of the catalyst section. These are extensions of the reactor container without obstacles, allowing streamwise realignment of the flow and enabling periodic boundary conditions. This strategy allows the application of a body force mimicking a pressure gradient, avoiding the manipulation of boundary pressures and improving the numerical stability~\cite{yang2023suitability}. At the fluid-solid boundaries, no-slip conditions are imposed. After steady state is reached, the flow Péclet number is $\Pen=Ud/D_m\approx 280$ in all simulations, $U$ being the mean flow velocity and $D_m$ the molecular diffusivity.
At this value of $\Pen$, chaotic mixing features are expected to emerge~\citep{lester2014control}.


The flow field is solved via a single-relaxation-time lattice Boltzmann equation:
\begin{eqnarray}
f_r (\boldsymbol{x} + \boldsymbol{c}_r, t+1) - f_r(\boldsymbol{x},t) =  - \tau_\nu^{-1} ( f_r(\boldsymbol{x},t)-f^{eq}_r (\boldsymbol{x},t)  )  + F_r \ ,
\label{eqlbm}
\end{eqnarray}
where $f_r(\mathbf{x},t)$ is the distribution function  at the position $\boldsymbol{x}=(x,y,z)$ and time $t$ along the $r$-th direction, $\boldsymbol{c}_r$ is the discrete velocity vector along the $r$-th direction, $\tau_\nu$ is the relaxation time  (proportional to fluid viscosity). With $f_r^{eq}$ we indicate the equilibrium distribution function along the $r$-th direction:
\begin{equation}
f^{eq}_r (\boldsymbol x,t) = w_r \rho\ \Bigg ( 1 +\frac{c_{rj} u_j(\boldsymbol x,t)}{c_s^2} + \frac{(c_{rj} u_j(\boldsymbol x,t))^2}{c_s^4} - \frac{u_j^2(\boldsymbol x,t)}{2c_s^2} \Bigg ) \ ,\label{eqf} 
\end{equation}
where $c_s$ is the speed of sound and $w_r$ the D3Q19 weight parameter of the three-dimensional lattice structure along the $r$-th direction. The fluid field is computed in each computational cell by integrating the hydrodynamic moments of the distribution functions. The steady state velocity vector $u_j(\boldsymbol{x})$, and density $\rho(\boldsymbol{x})$ are then obtained as:
\begin{eqnarray}
\rho(\boldsymbol x)  &=& \sum_r f_r(\boldsymbol x)   \\ 
\rho(\boldsymbol x)  {u}_j (\boldsymbol x) &=& \sum_r {c}_{rj} f_r(\boldsymbol x)  + \frac{1}{2}\bigg ( \frac{\Delta P}{L} \bigg )_j\ .
\end{eqnarray}
Here, ${\Delta P/}{L}$ is the applied pressure gradient that drives the fluid through the medium. This term also appears in the body force $F_r$ in eq.~\ref{eqlbm} as~\citep{guo2002discrete}:
\begin{equation}
F_r (\mathbf x,t)= \Bigg ( 1-\frac{1}{2\tau_\nu} \Bigg ) w_r \Bigg ( \frac{{c}_{rj}-{u}_j(\mathbf x,t)}{c_s^2} +\frac{{c}_{rj}{u}_j(\mathbf x,t)}{c_s^4}{c}_{rj} \Bigg ) \ \bigg ( \frac{\Delta P}{L} \bigg )_r \ .
\end{equation}
In the next step, the computed steady-state flow field is used to solve the transport of a passive scalar concentration $c$ via a second lattice population. The scalar is continuously injected at the inlet at concentration $c_0$ and may react at the catalyst-fluid interface.

The system is governed by the steady incompressible momentum equation:
\begin{equation}
u^*_j\dfrac{ \partial u^*_i(\boldsymbol{x})}{\partial x_j^*} = -\dfrac{\partial p^*}{\partial x^*_i}+\dfrac{\partial }{\partial x_j^*} \bigg (  \frac{1}{\Re} \dfrac{\partial u^*_j(\boldsymbol{x})}{\partial x_j^*} \bigg ) \ , \label{eqmom}
\end{equation}
and by the advection-diffusion-reaction equation:
\begin{equation}
\dfrac{ \partial c^*(\boldsymbol{x},t)}{\partial t^*} +   \dfrac{ \partial c^*(\boldsymbol{x},t) u^*_j(\boldsymbol{x})}{\partial x_j^*} = \dfrac{\partial }{\partial x_j^*} \bigg (  \frac{1}{\Pen} \dfrac{\partial c^*(\boldsymbol{x},t)}{\partial x_j^*} \bigg ) \ , \label{eqade}
\end{equation}
where $c^*(\boldsymbol{x},t)=c(\boldsymbol{x},t)/c_0$ is the dimensionless concentration, $u^*_j (\boldsymbol{x})= u_j (\boldsymbol{x})/U$ is the dimensionless velocity along the $j$-th direction, and $x_j^*=x_j/d$. We impose zero-flux boundary conditions at the container walls and outlet. The Reynolds number is set at $\Re=Ud/\nu=0.5<1$ and we confirmed that inertial effects are negligible by ensuring that the pore-scale flow statistics (probability distribution functions) are the same when computed at $\Re=0.005$ (not shown here). This result is in line with \cite{nissan2018inertial}, which reports non-negligible inertial effects from $\Re=10$.

At the catalyst surface, denoted by the inward-pointing normal direction $\lambda$, a first-order kinetic law is applied:
\begin{equation}
 - \frac{\partial c^*(\boldsymbol{x},t)}{\partial \lambda^*} \bigg |_{S} = \mbox{\textit{Da}}_\textit{I}  \ c^*(\boldsymbol{x},t) \big |_{S} \ , \label{eqads}     
\end{equation}
where $\mbox{\textit{Da}}_\textit{I}=k_s d/D_m$ is the pore-scale Damkhöler number, relating the characteristic velocity of surface reaction $k_s$ (in m/s units) to the pore-scale diffusive velocity $D_m/d$. Here, $\lambda^*=\lambda/d$ denotes the dimensionless interface normal. 

For the purpose of reactor design, we define a advection-based macroscopic Damköhler number:
\begin{equation}
\mbox{\textit{Da}}_\textit{II} = \dfrac{k_v L}{U} = \dfrac{k_s s_p (1-\phi)}{\phi}\dfrac{L}{U} , \label{eqdamm}
\end{equation}
which relates the macroscopic reaction rate $k_v = k_s s_p (1-\phi)/\phi$ to the mean residence time $L/U$, with $s_p=6/d$ the specific surface area of the catalyst particle and $\phi \approx 0.5$ the porosity.

We examine two cases for each geometric configuration: an inert case, with $\mbox{\textit{Da}}_\textit{II} = 0$, and a reactive case, with $\mbox{\textit{Da}}_\textit{II} \approx 2$. The latter ensures that the reactor operates under mass-controlled conditions and achieves optimized conversion~\citep{maggiolo2020solute}. The relevant dimensionless parameters are summarized in Table~\ref{tab1}, whereas in Table~\ref{tab2} we report the dimensional parameters relevant for packed bed reactor applications, such as MOST systems.

\begin{table}
  \begin{center}
\def~{\hphantom{0}}
  \begin{tabular}{lccccc}
      case & $R/d$  & $L/d$   &   $\mbox{\textit{Da}}_\textit{I}$ & $\mbox{\textit{Da}}_\textit{II}$ & \Pen \\[3pt]
     (A1) inert &  3.1     & 28 & 0 & 0 & 277 \\
     (A2) reactive &  3.1     & 28 &3.33 & 2 & 277 \\
     (B1) inert &  4.7     & 42 & 0 & 0 & 274 \\
     (B2) reactive &  4.7     & 42 &2.22 & 2 & 274 \\   
  \end{tabular}
  \caption{Simulated cases. The table reports the values of the reactor-to-particle size ratios, $R/d$ and $L/d$. The macroscopic Damköhler number is set to $\mbox{\textit{Da}}_\textit{II}=k_v L/U$, with values of 0 (inert case) and 2 (reactive case). In all simulations, the Péclet number is $\Pen = Ud/D_m \approx 280$, and the Reynolds number is $\textit{Re}=Ud/\nu \approx 0.5$.}
  \label{tab1}
  \end{center}
\end{table}

\begin{table}
  \begin{center}
\def~{\hphantom{0}}
  \begin{tabular}{lccccccc}
      case  & $d$ [$\mu$m] & $R$ [mm] & $L$ [cm] & flow rate [ml/h] & $D_m$ [m$^2$/s] & $k_v$ [1/s] \\[3pt]
     (A2)   &   833	            & 2.5 	        & 2.3 	 & 12 & 10$^{-9}$ & 3 10$^{-2}$ \\
     (B2)   &   533	            & 2.5 	        & 2.3 	 & 18 & 10$^{-9}$ & 3 10$^{-2}$ \\
     \end{tabular}
  \caption{Dimensional parameters for the two simulated reactors, cases A2 and B2 in Table~\ref{tab1}, relevant for, e.g., MOST applications. Values of particle diameters and reaction rates reflect those of typical catalyst employed, container dimensions and flow rates are comparable to those of small size miniature reactors~\citep{magson2024sust}. Molecular diffusivities correspond to typical values of hydrocarbons in liquids~\citep{price2000self}.}
  \label{tab2}
  \end{center}
\end{table}

\subsection{Lagrangian Particle Tracking and fluid deformation along streamlines~\label{sec:lag}}

We apply a Lagrangian Particle Tracking (LPT) algorithm to study the kinematics governing fluid deformation. Indeed, fluid deformation ultimately controls the spatial distribution of scalar quantities transported by the flow, i.e., their dynamic mixing. The stationary flow field obtained from the LBM simulations is used as input to the LPT algorithm, from which Lagrangian statistics are extracted. The governing equation of LPT,
\begin{equation}
\frac{\mathrm{d}\boldsymbol{x}(t,\boldsymbol{X})}{\mathrm{d} t} = \boldsymbol{u}[\boldsymbol{x}(t,\boldsymbol{X})] 
\end{equation}
is solved using a third-order Runge-Kutta method~\citep{maggiolo2016flow}, where $\boldsymbol{x}$ and $\boldsymbol{X}$ denote Eulerian and Lagrangian positions, respectively. We track 500 trajectories, randomly initialized at the inlet, and follow them until they reach, on average, half the length of the medium.

The extracted Lagrangian statistics are used in Section~\ref{sec:msd} to quantify macroscopic dispersion behavior, and in Section~\ref{sec:def} to analyze local fluid deformation. For the latter, we employ the procedure of \cite{lester2018fluid}, which involves constructing a Protean moving reference frame. For steady flows, the Protean frame is essentially a streamline frame. This transformation renders the velocity gradient tensor upper-triangular, simplifying the solution of the deformation tensor, and separating stretching from shear deformations.

The transformation is expressed as two successive reorientations, $\mathsfbi{Q}(t) = \mathsfbi{Q}_1(t)\mathsfbi{Q}_2(t)$. The first reorientation, $\mathsfbi{Q}_1$, aligns the velocity vector with the Protean frame. The second, $\mathsfbi{Q}_2$, applies an arbitrary rotation about the streamline direction. A specific choice of this rotation angle, $\alpha(t)$, ensures that the velocity gradient tensor becomes upper triangular. The angle $\alpha(t)$ is obtained from an ordinary differential equation whose coefficients depend on the velocity components $v_i(t)$, with $i=1,2,3$, and the velocity gradient tensor components $\epsilon_{ij}(t)$, evaluated at each time step along the Lagrangian trajectories.

The velocity components are directly available from the LPT solution for each tracer at each time step, while the velocity gradient tensor is computed via trilinear interpolation on the computational grid. Near solid surfaces, when a grid node required for interpolation lies outside the fluid domain, no-slip boundary conditions are enforced by mirroring the velocity from the nearest fluid node.

For further details on the Protean frame, we refer the reader to \cite{lester2018fluid}. A derivation of the velocity gradient tensor and a validation test case are provided in Appendix~\ref{sec:app}.

\section{Results~\label{sec:res}}

\subsection{Flow topology~\label{sec:top}}

We begin our analysis by characterizing pore-scale velocities and deformations. Figure~\ref{fig1}(b) shows the probability density functions (PDFs) of pore-scale velocities in the streamwise direction, $u_x(\boldsymbol{x})$, and in the transverse direction, $u_t(\boldsymbol{x})=(u_y^2(\boldsymbol{x})+u_z^2(\boldsymbol{x}))^{1/2}$. The results reveal that most high-velocity flow paths are longitudinally aligned. The difference between the streamwise and transverse velocity PDFs, which highlights this preferential streamwise alignment, depends only weakly on particle size across the investigated $R/d$ ratios. Moreover, the velocity PDFs are nearly identical for the two transverse confinement cases, $R/d=4.7$ and $R/d=3.1$. This indicates that wall-induced hydrodynamic effects are minimal, and the media can effectively be considered transversely unbounded (as confirmed by the wall-to-core fluid ratios $d/2/(2R-d)<0.1$~\citep{ziolkowska1988fluid}).

 \begin{figure}
  \centerline{\includegraphics[width=0.99\linewidth]{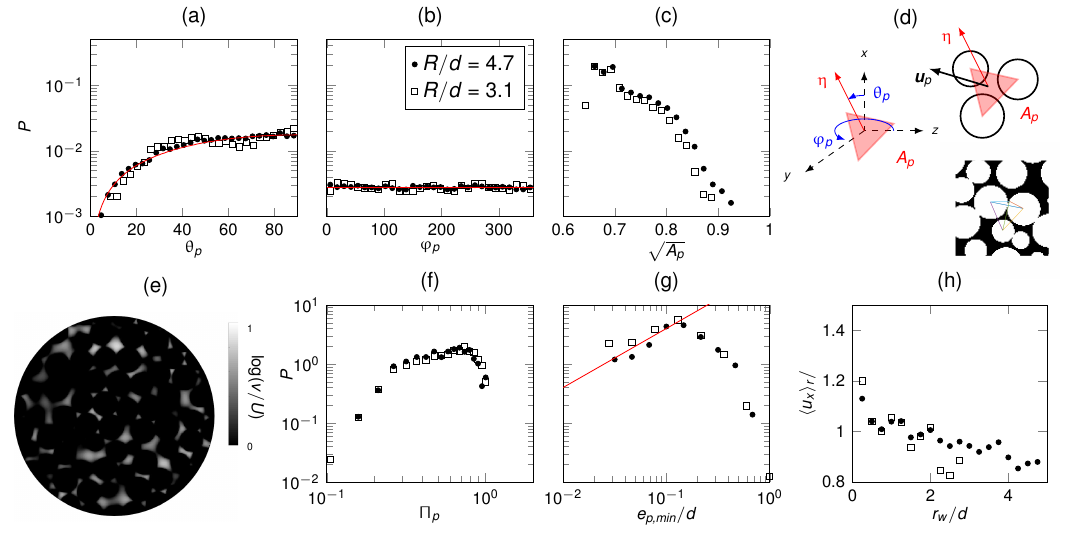}}
  \caption{PDFs of the polar ($\theta_p$) and azimuthal ($\varphi_p$) angles of the normal vector to pore constrictions $\boldsymbol{\eta}$ (panels a, b), and the PDF of constrictions' characteristic sizes $\sqrt{A_p}$ (panel c), computed via Delaunay triangulation. Panel (d) illustrates a triangular pore constriction (top), with the associated pore-scale velocity $\boldsymbol{u}_p$ and normal $\boldsymbol{\eta}$, and shows the segmented pore space with Delaunay triangulation (bottom). Red solid lines in (a) and (b) correspond to uniform distributions of $\cos \theta_p$ and $\varphi_p$. In (c), the sizes of pore constrictions exhibit a narrow distribution, with $\sqrt{A_p}/d \approx 0.5-1$. 
Panel (e) shows a cross-section of the flow field with the dimensionless pore-scale velocity magnitude $v/U$ displayed in logarithmic gray scale: this representation helps to assess the shape of the pore spaces, which are in large part non-circular. Panels (f) and (g) report the PDF of the ratio $\Pi_p$ between minimum and maximum axes of the ellipses inscribed in the pore spaces (f) and the PDF of the dimensionless minimum axis $e_{p,\textit{min}}/d$ (g). These panels confirm that pore spaces are non-circularly shaped, and that the probability of pore constrictions (small values of $e_{p,\textit{min}}$, panel (g), red line) qualitatively scales linearly with $e_{p,\textit{min}}$. In the last panel (h) we report the variation of the longitudinal velocity averaged over a cylindrical volume of length $L$ and diameter $R-r_w$, where $r_w$ is the distance from the packed bed container walls. Wall effects are clearly visible up to $r_w/d \approx 1/2$.
}
\label{fig2}
\end{figure}

At the pore scale, solid obstacles combined with no-slip boundary conditions generate viscous shear, which is important for deformation of scalar concentrations as well as for regulating surface reactions~\citep{aquino2023fluid}. The characteristic sizes of pore constrictions are identified using a three-dimensional Delaunay triangulation applied to the set of particle centers. This approach is well suited for spherical packings~\citep{ben2025pore}. For each triangle defined by three particle centers, we compute the pore constriction area $A_p$, its characteristic size $\ell_p = \sqrt{A_p}$, and the unit normal vector $\boldsymbol{\eta}$. The latter is not necessarily aligned with the positive streamwise direction. Its orientation relative to the Cartesian axes $(x,y,z)$ is quantified by the polar and azimuthal angles, $\theta_p$ and $\varphi_p$, as illustrated in figure~\ref{fig2}(d).

Three main descriptors are therefore used here to characterize the statistics of pore-scale topology: (i) pore constrictions $\ell_p$, (ii) streamwise alignment given by the polar angle $\theta_p$, and (iii) transverse orientation given by the azimuthal angle $\varphi_p$. The PDFs of $\theta_p$ and $\varphi_p$, reported in figure~\ref{fig2}(a,b), closely follow the theoretical uniform distributions of $\cos \theta_p$ and $\varphi_p$ (red lines), indicating no preferential orientation of pore-normal vectors. Pore constrictions (figure~\ref{fig2}c), by contrast, show a much narrower distribution, typically ranging from $\ell_p \sim 0.5d$ to $d$. This randomness in orientations is expected to influence how concentrations spread and mix, as will be discussed in Section~\ref{sec:agg}.

\subsection{Lagrangian fluid deformation~\label{sec:def}}

\begin{figure}
\centerline{\includegraphics[width=0.9\linewidth]{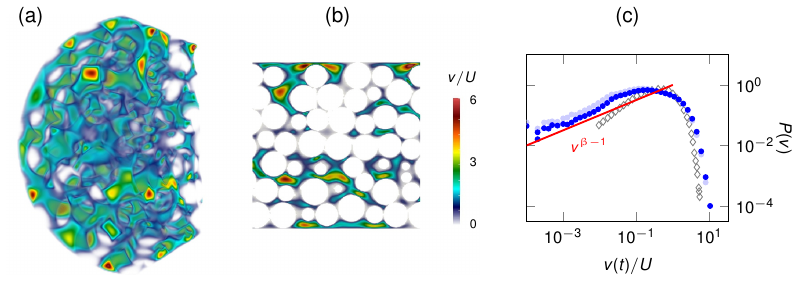}}
 \caption{Eulerian velocity magnitude fields $v/U$ on a volumetric slice (a) and on a cross-section in the $(x,z)$ plane (b).
(c) PDF of velocities $v/U$ sampled along Lagrangian trajectories. The exponent $\beta = 3/2$ is extracted from the low-velocity tail of the distribution using equation~\eqref{eqpdfvel} (red line). Dark and light blue circles correspond to the cases $R/d=4.7$ and $R/d=3.1$, respectively, while gray diamond markers report the data from~\cite{souzy2020velocity}.
}\label{fig3}
\end{figure}

Fluid deformation arises from different mechanisms. While shear flows induce algebraic deformation~\citep{souzy2018mixing}, stretching and folding processes cause exponential deformation of fluid elements in random media~\citep{heyman2020stretching}.
These mechanisms interplay to generate deformations as the product of a power law and exponential functions. To quantify the relative contributions of shear-induced deformation versus exponential deformation, we compute the deformation tensor $\mathsfbi{F}'(t)$ along Lagrangian trajectories in the Protean reference frame. Identifying the dominant deformation mechanisms is essential, as they govern solute mixing and porous medium's homogenization efficiency.

We first extract pore-scale velocity magnitudes $v=(u_i^2)^{1/2}$ along fluid trajectories, with $i=1,2,3$ indicating the three Protean directions. Figures~\ref{fig3}(a) and (b) display the Eulerian velocity magnitude fields, both in a 3D slice and on a symmetry plane. Sampling the dimensionless velocities $v/U$ along Lagrangian trajectories yields the PDF shown in figure~\ref{fig3}(c). From this distribution, we extract the power-law exponent $\beta$, assuming that low-velocity statistics follow the heavy-tailed form characteristic of heterogeneous media~\citep{berkowitz2006modeling}:

\begin{equation}
P(v<U) \propto v^{\beta-1}. \label{eqpdfvel}
\end{equation}
This analysis gives $\beta \approx 3/2$ (red line in figure~\ref{fig3}c). Interestingly, the low velocity distribution is not flat, as it would be if the pore cross-sections were circular~\citep{dentz2018mechanisms}. In circular cross-sections the PDF at low velocities follows the solution of the Poiseuille flow in a pipe, where regions of low velocity (near the pipe walls) are more probable than regions of high velocity. The high probability of low velocity regions is compensated by the low probability of low velocities in parabolic flow profiles, such as Poiseuille flows, resulting in a flat probability distribution of low velocities, i.e. $P(v<U) \propto v^0$. 
In non-circular cross-sections, the picture is different: in rectangular cross-sections the probability of low velocities increases with increasing velocity, because the probability of finding regions of low velocities idoes not continue increasing as the walls are approached (see e.g.~\cite{stewart2013note}; in the limit case of a two-dimensional flow the probability of low velocity regions is flat). 
Via image segmentation performed at several cross-sections (thresholding on $v/U$, see figure~\ref{fig2} (e)), we quantitatively discern and identify individual pore areas; we observe that pore spaces are largely non-circular and oblong, with a ratio between minimum and maximum inscribed axes $\Pi_p =e_{p,\textit{min}}/e_{p,\textit{max}} < 1$ and a distribution of pore-constrictions scaling qualitatively as $P(e_{p,\textit{min}})\propto e_{p,\textit{min}}$, see figure~\ref{fig3} panels (f) and (g). \cite{de2017prediction} showed that for a power-law distribution of pore throats in two dimension, the low velocity distribution scales as a power law with an exponent that is half of the exponent of the pore-throat distribution. Our findings are consistent with this result, as the low-velocity power-law exponent in equation~\eqref{eqpdfvel} satisfies $\beta - 1 = 1/2$. This suggests that the pore spaces are better approximated by quasi–two-dimensional geometries (with $e_{p,\textit{min}}$ denoting the 2D pore size), rather than by fully three-dimensional pores with circular cross-sections.

\begin{figure}
\centerline{\includegraphics[width=0.99\linewidth]{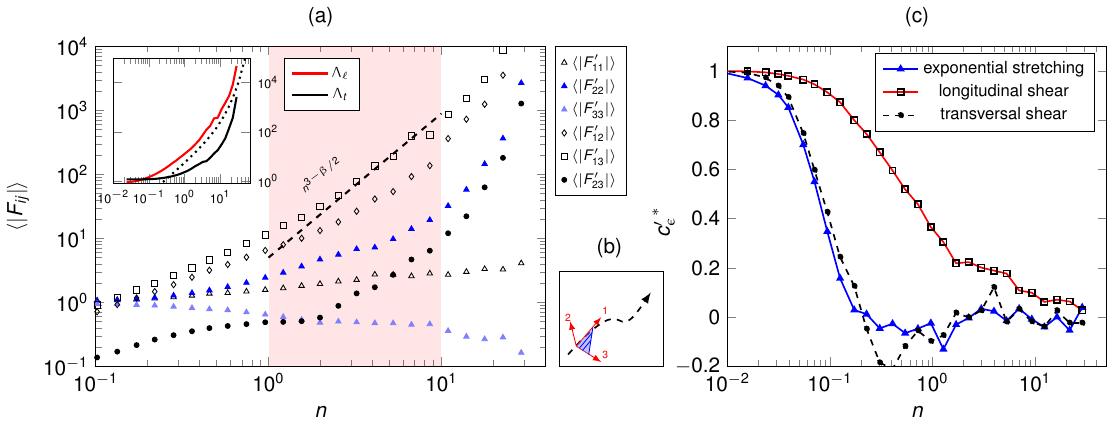}}
 \caption{
(a) Evolution of the deformation tensor components, averaged over the Lagrangian trajectories, for the case $R/d=4.7$. Longitudinal shear-induced components $F'_{12}$ and $F'_{13}$ dominate at early times ($n<10$), well captured by the prediction~\eqref{eqf12} (dashed lines). Inset: comparison between computed longitudinal and transverse deformations ($\Lambda_\ell$ as per equation~\eqref{eqalpha}, $\Lambda_t=\langle \sqrt{{F'_{22}}^2+{F'_{33}}^2+{F'_{23}}^2} \rangle$) and the cross-over prediction~\eqref{eqcross} (dotted lines). (b) Example of the Protean frame (red axes) orientation with respect to a Lagrangian trajectory (black dashed line); the blued shaded area schematically represents fluid shear deformation $F'_{13}$.
(c) Autocorrelation ${c'}^*_\epsilon$ of velocity gradient tensor components $\epsilon'_{ij}(t)$, showing long-lived correlations of shear rate components $\epsilon'_{12}$ and $\epsilon'_{13}$ (longitudinal shear) beyond a pore-scale travel time $n=1$, compared with rapidly decorrelating transversal shear rates $\epsilon'_{23}$ and exponential rates $\epsilon'_{11}$, $\epsilon'_{22}$, $\epsilon'_{33}$.
}\label{fig4}
\end{figure}

Next, we analyze the components of the deformation tensor, which reads as: 
\begin{equation}
\mathsfbi{F}'(t) = \Bigg ( \ \ 
\begin{matrix}
F'_{11}(t) & F'_{12}(t) & F'_{13}(t) \\
0 & F'_{22}(t) & F'_{23}(t) \\
0 & 0&  F'_{33}(t) \\
\end{matrix}
\ \ \Bigg )  \ .
\end{equation}
The off-diagonal terms $F'_{12}(t)$, $F'_{13}(t)$, and $F'_{23}(t)$, associated with deformation in the streamwise ($F'_{12}$, $F'_{13}$) and transverse ($F'_{23}$) directions, quantify shear-deformation at finite times and the combined effect of shear and stretching at later times. The diagonal terms $F'_{11}(t)$, $F'_{22}(t)$, and $F'_{33}(t)$ capture pure compression and stretching along the Protean axes induced by velocity intermittencies. Figure~\ref{fig4}a shows their time evolution: deformation is initially governed by shear, with $F'_{12}$ and $F'_{13}$ dominating up to $n \approx 10$ (red shaded region,$n=tU/d$ is the characteristic dimensionless time). At later times, exponential deformations (diagonal terms $F'_{22}$ and $F'_{33}$) become comparable in magnitude and are expected to dominate at longer times~\citep{lester2018fluid}. The early-time shear regime is particularly relevant in finite systems, where the macroscopic length is only a few particle diameters and reactions may occur before exponential stretching develops.

The mathematical framework of \cite{lester2018fluid} explains this behavior. At very early times, deformation evolves ballistically; prior to exponential stretching ($t<t_\epsilon$), shear dominates. Specifically, $F'_{13}$ follows a power law in $n$:
\begin{equation}
|F'_{13} (t<t_\epsilon)|  \approx \gamma^*_{2} n^{3 - \beta/2} \label{eqf12}
\end{equation}
where $\gamma^*_2$ is an effective shear rate in the $(1,3)$ plane of the Protean frame and $\beta$ is the velocity PDF exponent. Figure~\ref{fig4}a confirms that this model accurately captures the average deformation $\langle |F'_{13}(t)| \rangle$ (where $\langle \rangle$ indicates the averaging over the computed Lagrangian trajectories), with $\gamma^*_2 \approx 5$. This value is smaller than that reported for ordered bead packs
 ($\gamma^* \approx 40$)~\citep{aquino2023fluid}. This discrepancy is expected as ordered packs can trigger optimal mixing~\citep{turuban2018space,turuban2019chaotic}.
The evolution of $\langle |F'_{12}(t)| \rangle$ is similar to $\langle |F'_{13}(t)| \rangle$ but slightly slower, whereas $|F'_{22}(t)| \ll  |F'_{13}(t)|$, which is the main result of interest, showing that shear-induced deformation dominates at finite times. 
Algebraic functions (such as shear deformation) grows faster than an exponential ones (stretching) at short times. However the converse is true at long times. Hence the switch from shear to stretching at finite times is universal. Wether one deformation mode prevails on the other to determine mixing instead depends on the specific shear intensity, as we will discuss later in Section~\ref{sec:exp}.
Assuming $|F'_{12}| \sim |F'_{13}|$, the longitudinal growth of a material line at finite times is given by:
\begin{equation}
\Lambda_\ell(t<t_\epsilon) = \Big \langle \sqrt{{F'_{11}}^2(t)+{F'_{12}}^2(t)+{F'_{13}}^2(t)} \Big \rangle \approx \gamma^* n^\alpha, \label{eqalpha}
\end{equation}
with $\alpha = 3 - \beta/2 \approx 2.25$ and an effective shear rate in three dimensions $\gamma^* = (2{\gamma^*_2}^2)^{1/2} \approx 5\sqrt{2}$. This scaling will later be used to model concentration aggregation and mixing.


\begin{figure}
\centerline{\includegraphics[width=0.65\linewidth]{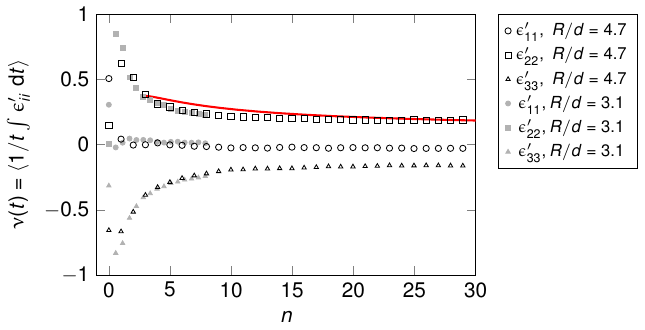}}
 \caption{Finite-time Lyapunov exponent $\langle \nu(t) \rangle$: comparison between numerical estimates (markers, obtained via the finite-time solution of equation~\eqref{eqlyapu}) and the prediction~\eqref{eqnu} with $\lambda=0.1$, $\beta=3/2$ (red line). Grey markers: case with stronger confinement, $R/d=3.1$.
}\label{fig5}
\end{figure}

The onset of exponential stretching occurs at the characteristic time $t_\epsilon = 1/\lambda$, where $\lambda$ is the infinite-time Lyapunov exponent:
\begin{equation}
\lambda = \big \langle \lim_{t \to \infty} \frac{1}{t} \int_0^t \epsilon'_{22}(t),\mathrm{d}t \big \rangle = -\big \langle \lim_{t \to \infty} \frac{1}{t} \int_0^t \epsilon'_{33}(t),\mathrm{d}t \big \rangle. \label{eqlyapu}
\end{equation}
Following~\cite{lester2018fluid}, the crossover between shear-induced and exponential deformation is described by
\begin{equation}
\Lambda_\ell(t) \approx \gamma^* n^{2 - \beta/2} \exp(\lambda t), \label{eqcross}
\end{equation}
while the finite-time Lyapunov exponent evolves as
\begin{equation}
\langle \nu(t) \rangle \approx \lambda + \bigg(2 - \frac{\beta}{2}\bigg)\frac{\ln(n)}{n}, \label{eqnu}
\end{equation}
which converges to $\lambda$ for $t,n \to \infty$. Numerically computing $\nu(t)$ via the solution of equation~\eqref{eqlyapu} at finite times and fitting with equation~\eqref{eqnu} yields $\lambda \approx 0.1$ (figure~\ref{fig5}, markers and red line). This value agrees well with the one computed for random stretching orientations~\citep{lester2016chaotic}. Predictions from equation~\eqref{eqcross} (dotted line) also match the observed longitudinal deformation across the crossover regime (figure~\ref{fig4} inset). Transverse deformation $\Lambda_t(t)$ should instead scale exponentially~\citep{lester2018fluid}, as confirmed by our simulations (inset of figure~\ref{fig4}a).
Combining \eqref{eqnu} with \eqref{eqf12} one can directly observe the dependence of the finite-time Lyapunov exponent on the dominant shear deformation $F'_{13}$:
\begin{equation}
\langle \nu(t) \rangle \approx \lambda + \frac{1}{n} \bigg [ \ln \big ( {\gamma^*_2}^{-1}\langle F'_{13}(n) \rangle \big ) - \ln(n) \bigg ] \ . \label{eqnu2}
\end{equation}

Finally, we note that the intermediate deformation regime is dominated by correlated shear events, which gradually lose correlation as the exponential regime is approached (figure~\ref{fig4}c). In particular, shear-induced deformations remain correlated well beyond a pore-scale travel time ($n=1$). The autocorrelation of velocity gradient components is computed as
\begin{equation}
{c'_{\epsilon}}^* = \frac{\langle (\epsilon'_{ij}(t)-\langle \epsilon'_{ij}(t)\rangle)(\epsilon'_{ij}(0)-\langle \epsilon'_{ij}(0)\rangle)\rangle}{\langle (\epsilon'_{ij}(t)-\langle \epsilon'_{ij}(t)\rangle)^2 \rangle^{1/2} \langle (\epsilon'_{ij}(0)-\langle \epsilon'_{ij}(0)\rangle)^2 \rangle^{1/2}},
\end{equation}
where $\epsilon'_{ij}(t)$ are components of the upper-triangular velocity gradient tensor. This correlation play a key role in controlling the randomness of scalar aggregation~\citep{duplat2010entanglement}, and it will be further discussed in Section~\ref{sec:agg}.

\subsection{Dispersion of tracers and evolution of the concentration front~\label{sec:msd}}

 \begin{figure}
  \centerline{\includegraphics[width=0.9\linewidth]{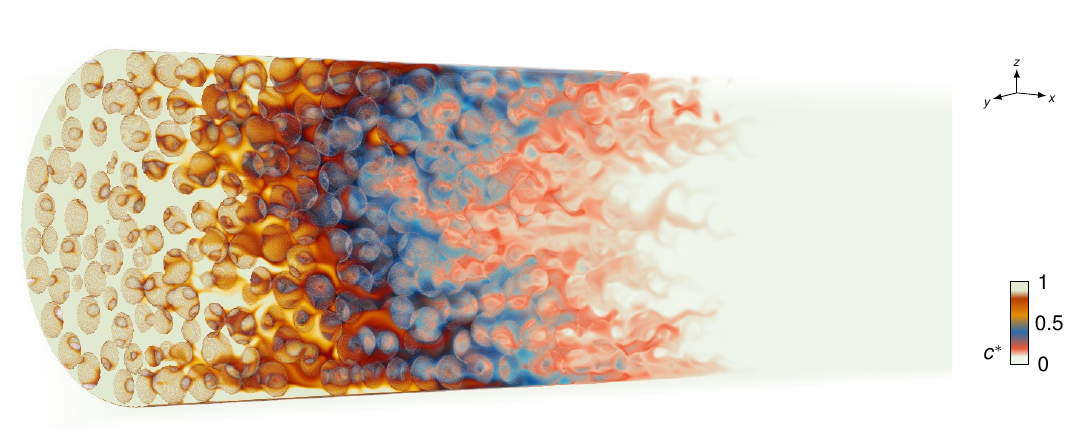}}
  \caption{
Snapshot of scalar transport in an inert porous reactor ($R/d=4.7$). The dimensionless concentration $c^*(\boldsymbol{x},t)=c(\boldsymbol{x},t)/c_0$ is injected at the left boundary at constant concentration $c_0$. Filaments (or sheets) of concentration are visible near the mean front position $c^*\approx 0.5$ (blue region).
}
\label{fig6}
\end{figure}

We now turn to the analysis of scalar dispersion and mixing.
We first analyze inert porous media, before extending the discussion to reactive systems. In this Section and Section~\ref{sec:mix} we analyze dispersion and mixing from Eulerian data of pore-scale simulations, whereas in Sections~\ref{sec:sp},~\ref{sec:agg} and \ref{sec:exp} we interpret the results with mixing models based on Lagrangian deformation.

\begin{figure}
  \centerline{\includegraphics[width=0.99\linewidth]{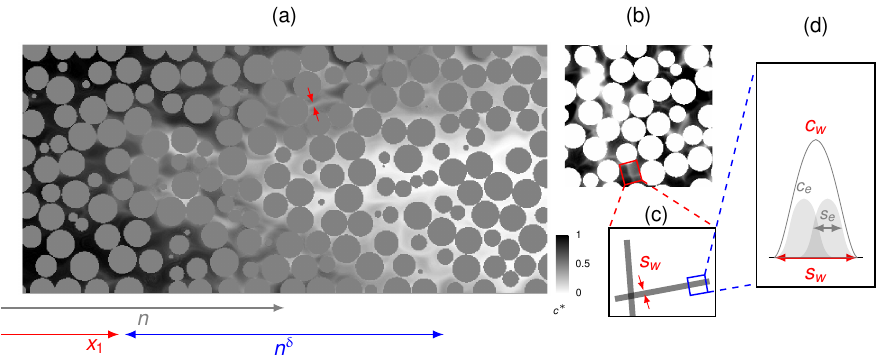}}
  \caption{(a) Snapshot showing the longitudinal extension $n^\delta$ of the dispersive volume, centered at the mean front position $Ut/d=n$. Low-concentration regions ({\em whites}) form in particle wakes (red arrows). (b) Cross-section showing the sheet-like, three-dimensional structure of white regions (red arrows in panel (a)). (c) Magnification of the two overlapping sheets in the red rectangle of panel (b). The white sheets, of width $s_w$, are shown here in gray. (d) The observed sheets are bundles of several elementary concentration sheets with maxima $c_e$ and thickness $s_e$. Bundles have mean thickness $s_w$ and concentration $c_w$.}
\label{fig7}
\end{figure}

When a scalar is injected into the porous reactor, flow deformation strongly influences the distribution of concentration. This effect is especially evident near the mean concentration front, i.e. close to the isosurface $c^*\approx 0.5$ (figure~\ref{fig6}). Here, filaments of concentration emerge along high-velocity flow paths, which correspond to regions of strong deformation. More specifically, concentration sheets are formed and stretched preferentially along the dominant flow direction. The same phenomenon occurs for the complementary concentration field $c_w=1-c^*$, which we refer to as the {\em white} concentration elements (in contrast to the {\em black} high-concentration regions). The most prominent white regions, corresponding to low concentration, appear in the wakes of catalyst particles (figure~\ref{fig7}a). In cross-section, these regions resemble two-dimensional sheets (figure~\ref{fig7}b), which occasionally overlap transversely (red arrows).

The mean concentration front advances as $n = Ut/d$, which also represents the mean number of encountered solid obstacles. White concentration elements accumulate near the mean front, elongating over time and increasing in number from the rear to the head. This leads to a progressive transformation of the initial step profile ($c^*=1$, $c_w=0$ at the back; $c^*=0$, $c_w=1$ at the head) into a smoother longitudinal profile. The smoothing is the result of longitudinal dispersion, which spreads concentration elements along the streamwise direction.

\begin{figure}
  \centerline{\includegraphics[width=0.99\linewidth]{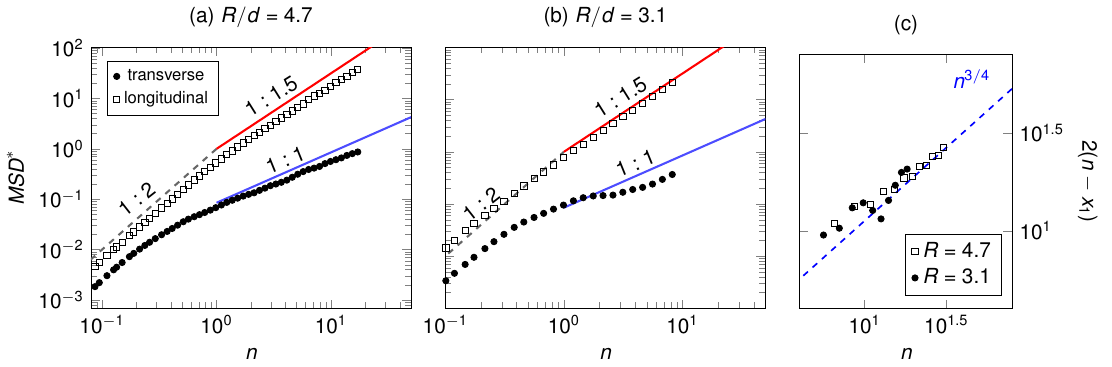}}
  \caption{Mean square displacement (MSD) computed from Lagrangian trajectories (equation~\eqref{eqmsd}). The longitudinal MSD shows superdiffusive behavior for (a) $R/d=4.7$ and (b) $R/d=3.1$. (c) Comparison between dispersive lengths computed: (i) from Lagrangian statistics, $\sqrt{\textit{MSD}_\ell}$ (blue dashed line), and (ii) from the Eulerian field, $2(n-x_1)$ (markers), where $n$ and $x_1$ are the mean and lowermost front positions (see figure~\ref{fig7}a).}
\label{fig8}
\end{figure}

To quantify the evolution of the concentration front, we compute longitudinal dispersion from the mean square displacement (MSD) of Lagrangian tracers:
\begin{eqnarray}
\textit{MSD}_\ell &=& \dfrac{\langle [ (x_k(t)-x_k(0))-\langle x_k(t)-x_k(0)\rangle ]^2 \rangle}{d^2}, \\ 
\textit{MSD}_t &=& \dfrac{1}{2d^2} \Big( {\langle [ (y_k(t)-y_k(0))-\langle y_k(t)-y_k(0)\rangle ]^2 \rangle} \nonumber \\ 
&& + {\langle [ (z_k(t)-z_k(0))-\langle z_k(t)-z_k(0)\rangle ]^2 \rangle} \Big), \label{eqmsd}
\end{eqnarray}
where $\textit{MSD}_\ell$ and $\textit{MSD}_t$ denote the longitudinal and transverse MSDs, and $(x_k(t),y_k(t),z_k(t))$ are the positions of Lagrangian tracers. After the characteristic flow time $n=1$, the longitudinal MSD exhibits superdiffusive scaling, while the transverse MSD is diffusive (figure~\ref{fig8}).  

Longitudinal dispersion spreads tracers along the mean flow direction, producing a dispersive volume whose streamwise length can be also estimated from the Eulerian concentration field. We identify the lowermost position of this volume, $x_1$, as the first cross-section downstream of which the mean value of $c_w$ remains strictly positive. We then compare the measure of the longitudinal dispersive length estimated from the Eulerian concentration field as $2(n-x_1)$ (see the schematic in figure~\ref{fig7}a), with the Lagrangian evaluation $\sqrt{\textit{MSD}_\ell} \propto n^\delta$, where $\delta={3/4}$ is extracted from the superdiffusive trend shown in figure~\ref{fig8}. The two measures agree well (figure~\ref{fig8}c), showing the consistency among simulated Eulerian and Lagrangian data.

The longitudinal dispersion coefficient thus scales with the mean velocity, $D_\ell \approx Ud$, and the concentration front spreads within a volume whose length increases quasi-linearly in time as $n^\delta$ with $\delta \approx 0.75$. At longer times, we expect this superdiffusive regime to transition to Fickian spreading, consistent with observations in anisotropic~\citep{maggiolo2016flow} and highly heterogeneous media~\citep{souzy2020velocity}. 
The linear $D_\ell - U$ scaling at $\Pen=O(10^2)$ is qualitatively in line with previous works, where the longitudinal dispersion coefficient is expected to scale as $D_\ell  \propto \Pen^{1.19}$ up to $\Pen \approx 400$ and transitions to linear scaling at higher $\Pen$~\citep{bijeljic2004pore}. Transverse dispersion should show a more robust relationship, scaling as $D_t \propto \Pen$~\citep{bijeljic2007pore}; in random packed beds, this behavior has been experimentally observed at $\Pen \gtrapprox 10^3$~\citep{scheven2013pore}.

It should be noted that, as an alternative method, longitudinal and transverse dispersion of tracers can be also computed from the velocity distributions via continuous-time random walk theory, which provides relations between Eulerian and Lagrangian velocities for sampling along streamlines. For further reading the reader is referred to~\cite{dentz2016continuous}.

\subsection{Evolution of pore-scale mixing~\label{sec:mix}}

To characterize mixing, we use the concentration variance as a metric for mixing dynamics. At the pore scale, other common measures include the dilution index~\citep{chiogna2012mixing}. The concentration variance is capable of capturing small-scale variability, given sufficient resolution of observation. This can be computed either over a fixed or moving longitudinal region (``fluid-supported") or within a subset of that region where the concentration exceeds a threshold (``concentration-supported")~\citep{lester2016chaotic}. In this work, we adopt the fluid-supported variance as the primary mixing metric.

We focus on a subvolume near the mean concentration front, defined by $c^*=0.5$, that travels along the reactor. In practice, we select a discrete disc of thickness $md$ (with $m$ a constant, chosen between 4 and 6, and $d$ the particle size) such that the volumetric average concentration is $\mu_v = 0.5$. This choice reduces the effect of temporal fluctuations in the mean when computing the variance. The subvolume travels approximately with $n = Ut/d$. To avoid pronounced wall effects (see also panel (h) of figure~\ref{fig2}), we restrict the area of the disc $A$ to a circle of radius $R-d/2$.

\begin{figure}
  \centerline{\includegraphics[width=0.99\linewidth]{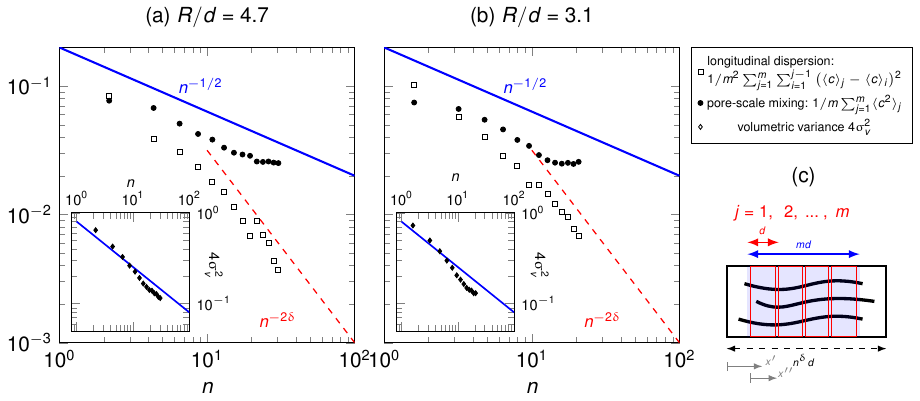}}
  \caption{Contributions to the concentration variance $\sigma_v^2$ (equation~\eqref{eqvar}) for (a) $R/d=4.7$ and (b) $R/d=3.1$ (averaged over two realizations). Longitudinal dispersion (empty squares) decreases in time as $n^{-2\delta}$, while pore-scale mixing (filled circles) decays more slowly $\sim n^{-1/2}$ (solid blue line, equation~\eqref{eqoverlap}) and dominates the total variance at long times (insets). Panel (c) illustrates the subdivision of the dispersive volume into 3D discs used to compute equation~\eqref{eqvar}.}
\label{fig9}
\end{figure}

For clarity, we define a shifted streamwise position $x'' = (n-m/2)d$ (see figure~\ref{fig9}, right panel), 
which is the position within the subvolume of length $m$ centered at the mean front position $n$. The volumetric variance in the subvolume is then calculated as:
\begin{equation}
\sigma^2_v(t)  = \dfrac{1}{Amd} \int_{0}^{md}  \int_A {c}^2(\boldsymbol{x}_f,t) \mathrm{d}A \ \mathrm{d}x'' \ - \mu_v^2(t) \ , \label{eqvardef}
\end{equation}
where $c(\boldsymbol{x}_f,t)$ is the concentration at fluid location $\boldsymbol{x}_f$ and $\mu_v$ is its volumetric mean. To separate longitudinal dispersion from pore-scale mixing, we introduce the spatial averaging operator $\langle \ \rangle_j$, applied to a 3D disc of thickness $d$ and area $A$, centered at $x''/d = (2j-1)/2$, $j=1,...,m$. Then the first and second moments in equation~\eqref{eqvardef} become:
\begin{eqnarray}
\mu_v =  \dfrac{1}{m} \sum_{j=1}^m \dfrac{1}{Ad} \int_{(j-1)d}^{jd} \int_A c(\boldsymbol{x}_f,t)  \mathrm{d}A \ \mathrm{d}x''  =  \dfrac{1}{m} \sum_{j=1}^m \langle c \rangle_j \nonumber \\  
 \dfrac{1}{Amd} \int_0^{md} \int_A {c}^2(\boldsymbol{x}_f,t) \mathrm{d}A \ \mathrm{d}x''  =  \dfrac{1}{m} \sum_{j=1}^m \langle c^2 \rangle_j \ . \label{eqvarmean}
\end{eqnarray}

Combining equations~\eqref{eqvardef} and~\eqref{eqvarmean}, the concentration variance can be rewritten as:
\begin{equation}
\sigma^2_v(t)  = \dfrac{1}{m^2} \sum_{j=1}^m \sum_{i=1}^{j-1} \big (  \langle c \rangle_j -  \langle c \rangle_i \big )^2 + \dfrac{1}{m} \sum_{j=1}^m \langle {c'}^2 \rangle_j \ , \label{eqvar}
\end{equation}
with $c' = c - \langle c \rangle_j$. The first term on the right-hand side quantifies one-dimensional longitudinal dispersion, measuring differences between cross-sectional average concentrations. It scales with the squared longitudinal concentration gradient, $(c_0^*/L_\ell)^2$, where $L_\ell^2 \sim \textit{MSD}_\ell \propto n^{2\delta}$ (Section~\ref{sec:msd}), and thus scales as $n^{-2\delta} \approx n^{-1.5}$ for $\delta \sim 0.75$. 

The second term captures transverse variations, measuring pore-scale differences and mixing, with longitudinal pore-to-pore differences removed by computing $\langle \cdot \rangle_j$ over discs of thickness $\lessapprox d$, the characteristic length scale of velocity fluctuations.

Figure~\ref{fig9} shows the evolution of the two contributions to the variance. Longitudinal dispersion reduces concentration differences along the flow direction, scaling as $n^{-2\delta}$ (empty squares). Pore-scale mixing decays more slowly (filled circles) and dominates the mixing evolution in the subvolume. This behavior is important at finite times and will be analyzed further in the next sections.

\subsection{Lamellar model in shear flow~\label{sec:sp}}

We observe a marked elongation of concentration regions along the streamwise direction and now ask how they form and arrange in space. To address this, we examine the distribution of {\em white} concentrations. These low-concentration regions can be regarded as bundles, or aggregates, of several elementary concentration sheets originating from particle wakes. We denote by $c_e$ the maximum concentration of the elementary sheets, and by $c_w$ the maximum concentration within a bundle of mean thickness $s_w$. A sketch of bundle formation is shown in figure~\ref{fig7}, panels (c) and (d).

Each elementary sheet originates in particle wakes and is subsequently deformed by the flow. We can make use of a lamellar model~\citep{villermaux2019mixing,dentz2023mixing} to model their evolution under the dominant deformation mechanism at finite times, that is, shear flow. Shear-induced elongation scales as $\Lambda_\ell \propto n^\alpha$, with $\alpha$ extracted from the Lagrangian kinematics in~\eqref{eqalpha}. The maximum concentration of each sheet after a short mixing time $t_m$ decreases as $c_e = c_0/(s_e n^\alpha)$, with $s_e\propto n^{1/2}$ the transverse size. Since $\alpha \approx 2.25$ exceeds the exponent $\delta \approx 0.75$ of the dispersive volume growth ($n^\alpha > n^\delta$), sheets overlap because of shear deformation at a rate $n^{\alpha-\delta+1/2}$, yielding an effective concentration $c_e n^{\alpha-\delta} s_e$. Overlap is further enforced by their increasing density: since they elongate along the flow direction, the number of elementary sheets per cross-section increases as $n_e = x' n_p$, from dilute ($x'\sim 0$) to dense regions ($x'\sim n$).  We emphasize that the density of concentration sheets grows in time since more sheets reach a given longitudinal position.

\begin{figure}
  \centerline{\includegraphics[width=0.99\linewidth]{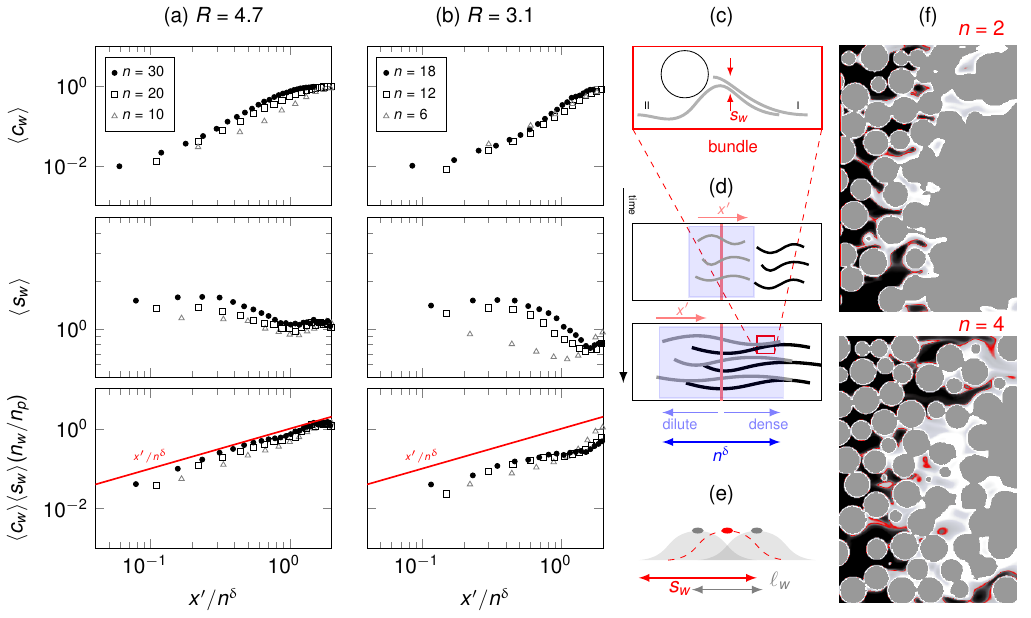}}
  \caption{(a,b) Upper panels: sample mean $\langle c_{w} \rangle$ per cross-section of the maximum concentrations in white bundles, developing longitudinally along the dispersive volume at different times $n$, for $R=4.7$ (a) and $R=3.1$ (b). Middle panels: mean lateral bundle size $s_w$, which grows with $n$ in the dilute region (blue lines). Lower panels: verification of the mass-conservation equation~\eqref{eqmass} (red lines).  
  (c) Sketch of elementary sheets of low concentration $c_e$, elongated longitudinally by the flow (I, II), and overlapping to form bundles of concentration $c_w$ and thickness $s_w$. 
  (d) Bundles are embedded in a dispersive volume of length $\sim n^\delta$ (blue shaded region). Dilute and dense regions of white concentrations are identified upstream and downstream of the mean front position (red vertical line).  
  (e) Bundle overlap occurs when the bundle width $s_w$ exceeds their spacing $\ell_w$, producing new maxima (red dot), with $n_w/n_p \sim s_w/\ell_w$.
In the rightmost panel (f) two snapshots of concentration fields show the emergence of white concentration elementary sheets, forming bundles at downstream positions (top), and later growing in number and merging (bottom).}
\label{fig10}
\end{figure}

Aggregation of elementary sheets thus produces bundles (figure~\ref{fig10}c and f), and mass conservation at each $x'$ requires:
\begin{equation}
\langle c_{w} \rangle \langle s_w \rangle \dfrac{n_w}{n_p} = c_e n^{\alpha-\delta} s_e \dfrac{n_e}{n_p} = \dfrac{x'}{n^\delta} \ , \label{eqmass}
\end{equation}
with $c_0=1$ and the symbol $\langle \rangle$ here indicating the sample average per cross section.

We verify this model numerically, from the pore-scale concentration field. We track the maxima $c_w$ at different cross-sections located within the interval $x'_1=[x_1 : x_1+ (n d))^\delta]$, corresponding to the longitudinal dispersive length centered at the mean front position $Ut/d=n$ (figure~\ref{fig7}a). The top panels of figure~\ref{fig10} show the sample mean of the maxima $\langle c_w \rangle$ at different dimensionless times $n$, increasing with the longitudinal position $x' = x'_1-x_1$ within the dispersive volume.

In the dilute region, bundles consist of few sheets and weak aggregation, leading to a power-law decrease of $\langle c_w \rangle$ (figure~\ref{fig10}, upper panels) as a result of elongation and diffusion. In the dense region, aggregation dominates and overlap drives $\langle c_w \rangle \to 1$. 

Diffusion dilutes the white bundles transversely. The mean lateral spreading $\langle s_w \rangle$ can be inferred from the number of maxima per pore $n_w/n_p$ and their mean separation distance $\langle \ell_w \rangle$, where $n_p=(2R/d)^2$ is the mean number of pores per cross-section. Local maxima arise at intersections of different bundles, and their number per pore is expected to increase with bundle overlap, estimated as $\sim \langle s_w \rangle/\langle \ell_w \rangle$. This leads to
\begin{equation}
\langle s_w \rangle \sim \langle \ell_w \rangle \dfrac{n_w}{n_p} \propto n^{1/2} \ , \label{eqsw}
\end{equation}
implying diffusive growth in time if the elementary sheets expand diffusively as $s_e \propto t^{1/2}$. (In equation~\eqref{eqsw}, $\ell_w$ and $s_w$ are normalized by the particle size $d$.) From the middle panels of figure~\ref{fig10} we observe that $\langle s_w \rangle$ increases with $n$ at fixed $x'$, supporting the interpretation of diffusive growth of elementary sheets and subsequent bundle overlap.
Finally, in the lowermost panels of figure~\ref{fig10}a and b, we also verify equation~\eqref{eqmass}.

Given this satisfactory agreement between the simulations and the lamellar model of deformation in shear flow, we next analyze the mixing dynamics in more detail and provide a more precise prediction through lamellar aggregation. In particular, we focus on the second moments of the concentration maxima, which provide an accurate measure of pore-scale mixing.

\subsection{Aggregation model under shear-induced deformation~\label{sec:agg}}

We adopt the shear-flow deformation model~\citep{villermaux2019mixing,souzy2018mixing} to describe the dynamics of mixing at finite times. We noted earlier that this model is consistent with the first moments of the white concentration sheets: it is the shear flow together with diffusion that induces the formation of white concentration sheets at the particle wakes and the successive blending at finite times.

We here briefly recall the shear-flow deformation model. At very early times, advection dominates over diffusion, and flow-induced elongation compresses concentration sheets along the transverse direction. As shear-induced deformation is dominant (see Section~\ref{sec:def}), the transverse thickness of these white sheets, $s_e(t)$, decreases according to:
\begin{equation}
s^{(1)}_e (t) \approx s_0 (\gamma t)^{-\alpha} \ , \label{eqs1}
\end{equation}
where $s_0$ and $s_0^*=s_0/d$ denote the dimensional and dimensionless initial thickness, respectively.

Applying Ranz's transformation~\citep{ranz1979applications} and solving the associated advection-diffusion equation in the transformed Lagrangian frame, one can estimate the characteristic mixing time $t_m$, at which diffusion ($D/s^2(t)$) begins to counterbalance advection-induced compression ($-(1/s_e)(\mathrm{d}s_e/\mathrm{d}t)$):
\begin{equation}
t_m \sim \dfrac{1}{\gamma}(\gamma^* {s^*_0}^2\textit{Pe})^{1/(2\alpha+1)} \ , \label{eqmixing}
\end{equation}
where $\gamma = \gamma^* U/d$ is the dimensional shear-induced deformation in three dimensions, as defined in Section~\ref{sec:def}.
Using the measured $\gamma^*= 5 \sqrt{2}$ and $\alpha=2.25$, we obtain $t_m \approx d/(3U)$. 

\begin{figure}
  \centerline{\includegraphics[width=0.99\linewidth]{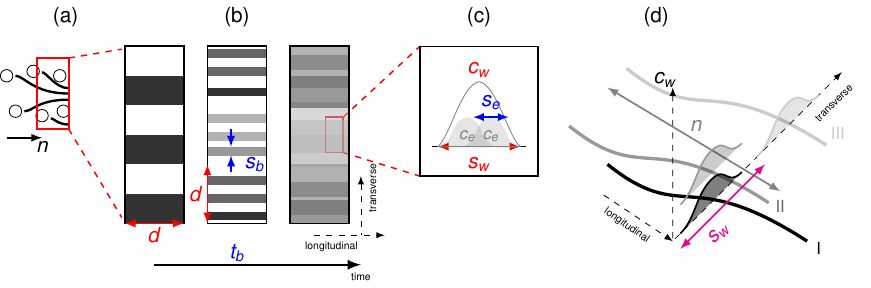}}
  \caption{(a) At initial times  low concentration sheets emerge from particle wakes because of shear flow. Their initial transverse size is $s_0=d/2$, so that the total concentration $c^*=0.5$. (b) Schematic illustration of the pore-scale mixing: at the mixing time $t_m$ the sheets has reached their minimum thickness $s_b$; after the mixing time $t_m$ diffusion merges them, to form bundles of maximum concentration $c_w$ and thickness $s_w$ (c). In (d) the two modes of overlap are sketched: the concentration sheet I overlap longitudinally with sheet II because of streamwise elongation, while transverse overlap between II and III is sustained by diffusion.}
\label{fig11}
\end{figure}

At this mixing time $t_m$, elongation and compression yield white concentration sheets with transverse thickness:
\begin{equation}
s^{(1)}_b \approx s_0 (\gamma^* {s_0^*}^2 \textit{Pe})^{-\alpha/(2\alpha+1)} \ , \label{eqsb1}
\end{equation}
which subsequently broadens diffusively as $s_e^{(1)} \sim \sqrt{2Dt}$. These concentration sheets are predominantly elongated along streamlines and thus mainly aligned with the streamwise direction (figure~\ref{fig11}, panel a). 

Up to $t_m$, in the vicinity of the front, elongation brings together an increasing number of concentration sheets originating from different streamwise positions, while compression prevents their overlap. Beyond $t_m$, transverse diffusion allows these elements to broaden as $t^{1/2}$, and their maximum concentration decays according to:
\begin{equation}
c_e \approx \dfrac{c_0 \ s_0}{(\gamma t)^\alpha \sqrt{2Dt}}\propto n^{-(\alpha+1/2)} \ . \label{eqce}
\end{equation} 
At this stage, the elements are free to overlap due to the combined effects of shear-induced deformation and transverse diffusion, as schematically illustrated in figure~\ref{fig11}. The key positive effect of early-time deformation is that it enhances mixing: by compressing concentration elements transversely, advection brings them closer together, allowing diffusion to merge them more efficiently.

We thus define an early-time aggregation model based on shear-induced elongation, the dominant mechanism of deformation at finite times, while exponential stretching dominates at longer times.
Under shear-induced deformation, aggregation of white concentration sheets becomes significant at times $t > t_m$ due to two main mechanisms: (i) the mismatch between the longitudinal growth of the sheets and the dispersive volume, occurring at a rate $n^{\alpha-\delta}$, and (ii) the increase in sheet density and reduction in their physical separation, with rate $N(t)$. The increase in physical proximity drives overlap as more longitudinally elongated concentration elements occupy the same pore spaces (Figure~\ref{fig6}), and transverse diffusion broadens their effective thickness.

The total number of overlaps per pore driven by proximity is estimated as:
\begin{equation}
N(t> t_m)\approx (\gamma t)^\alpha \sqrt{2Dt}/d \propto n^{\alpha+1/2} \ , \label{eqbign}
\end{equation}
which dominates aggregation dynamics since $N > n^{\alpha-\delta}$. Consequently, the average concentration of aggregates, $c_e N$, remains approximately constant near the front. Mass conservation is also reflected in the first moment of the white bundle maxima, $\langle c_w(x=n) \rangle$, as shown in Figure~\ref{fig10}. Nevertheless, this constancy does not apply to the variance.

Given that the mean concentration near the front is constant, the random aggregation model of~\cite{duplat2008mixing} can be applied to estimate the variance decay, based on the count of overlapping events among concentration sheets. According to this model, aggregation among decorrelated concentration sheets can be modeled via the random addition of independent and identically distributed variables, or via the sum of the means and variances when they are correlated~\citep{duplat2010entanglement}. In the former case, one makes use of the central limit theorem, based on the assumption that concentrations within isolated sheets or bundles are identically distributed and picked up randomly when performing aggregation (see also~\cite{weiss2007transport}).

The number of sheets per pore at $t_m$ is the consequence of their early-time elongation beyond a pore space:
\begin{equation}
n_b = (\gamma t_m)^\alpha \ ,
\end{equation}
which, combined with equations~\eqref{eqmixing} and~\eqref{eqsb1}, defines the  average dimensionless spacing between sheets:
\begin{equation}
\ell_b = 1/n_b = s_b/s_0 \ . \label{eqdistance}
\end{equation} 

After $t_m$, sheet overlap occurs via two mechanisms: the persistence of longitudinal, shear-induced elongation ($N_\ell$), and transverse diffusion ($N_t$), so that the total number of overlaps is $N = N_\ell N_t$, as expressed in~\eqref{eqbign}:
\begin{eqnarray}
&&N_\ell(t>t_m) = (\gamma t)^\alpha \ell_b \ , \nonumber \\
&&N_t(t>t_m) = s_e/(\ell_b d) \approx \sqrt{2Dt}/(d\ell_b) \ . \label{eqnn}
\end{eqnarray}

The variance of a single, isolated concentration element is proportional to its squared maximum concentration~\citep{meunier2010diffusive}:
\begin{equation}
\sigma^2_e(t) \approx c_e^2 \propto n^{-2\alpha-1} \ . \label{eqsingle}
\end{equation}
Longitudinal overlap aggregates these elements into bundles (Figure~\ref{fig10}, panels c and f). Because shear-induced deformation is correlated over lengths larger than the pore scale $d$ (Section~\ref{sec:def}), the bundle variance can be approximated by the correlated sum of $N_\ell$ individual variances:
\begin{equation}
\sigma_w^2(t) \approx c_{w,\textit{max}}^2 = N_\ell^2 \sigma_e^2 \propto n^{-1} \ . \label{eqbund}
\end{equation}

In contrast, transverse overlap occurs randomly across the cross-section. Transverse overlap yields values of concentrations that are the sum of different concentrations belonging to bundles interpenetrating with random orientations in the cross sectional space. This random aggregation rule is justified by the random orientation of azimuthal pore throats, observed in Section~\ref{sec:top}, and reported in Figure~\ref{fig2}.
Following this reasoning, the maximum concentrations near the front result from the random addition of $N_t$ values of concentration distributed according to the characteristic PDF of the bundle. Using equations~\eqref{eqce} and~\eqref{eqnn}, we obtain:
\begin{equation}
\sigma^2(t) \approx N_t \sigma_w^2 =  N_t N_\ell^2 c_e^2 = B n^{-1/2} \ , \label{eqoverlap}
\end{equation}
where $N_t(t)$ quantifies the evolution of the number of diffusive overlap and $B$ is a constant depending on the initial conditions $c_0, s_0$, the shear-induced deformation rate $\gamma^*$ and the Péclet number as:
\begin{equation}
B = \dfrac{c_0^2 s_0}{\sqrt{2}} \Bigg ( \dfrac{ {s_0^*}^2\Pen}{{\gamma^*}^{2\alpha}} \Bigg ) ^{1/(4\alpha+2)} \ . \label{eqacon}
\end{equation}
To better highlight the \Pen-dependence one can rewrite eq.~\eqref{eqoverlap} as:
\begin{equation}
\sigma^2(t) \propto ({\gamma^* \Pen})^{-\alpha/(2\alpha+1)} t^{-1/2}\ , \label{eqoverlap2}
\end{equation}
which, in the case of linear shear $\alpha=1$, yields the same scaling as that proposed for the variance of travel times of species reaching solid surfaces in porous media, as predicted by chaotic restart models~\citep{aquino2023fluid}. The present analysis suggests that one can expand such models to account for non-linear shear and  $\alpha \ne 1$.

Numerical results confirm that the variance near the front is well approximated by the variance of bundle maxima, $\sigma^2(t) \sim \langle c_w^2\rangle$ (Figure~\ref{fig12}, panel c). Evaluating equation~\eqref{eqacon} with the value of $\gamma^*$ extracted from Lagrangian deformations (Section~\ref{sec:def}) yields $B \approx 0.2$, which, when substituted into equation~\eqref{eqoverlap}, provides predictions in good agreement with pore-scale mixing Eulerian data (Figure~\ref{fig9}, blue solid line vs filled marks, large panels and insets). This confirms that mixing at finite times is dominated by shear-induced deformation.

The proposed aggregation model, combining correlated longitudinal and random transverse overlaps, predicts a slower decay of concentration variance than fully three-dimensional random aggregation, where variance scales as $n^{-3/2}$. However, early-time elongation and compression accelerate transverse overlap compared to pure diffusion, leading to variance decay that scales with the mean velocity rather than the molecular diffusion coefficient $D$, consistent with $n \propto U$.

\subsection{Shear-induced deformation versus exponential stretching~\label{sec:exp}}

At longer times, exponential stretching begins to dominate. However, slow shear-induced mixing continues to govern the overall mixing process beyond the characteristic onset time of exponential stretching, i.e., $n \sim 1/\lambda \approx 10$.

During exponential stretching, concentration elements must reach a stable minimum thickness that is independent of the initial concentration~\cite{villermaux2019mixing}:
\begin{equation}
s^{(2)}_b \approx d (\lambda \textit{Pe})^{-1/2} \ . \label{eqsb2}
\end{equation}
By inserting the values of the dimensionless shear-induced deformation rate $\gamma^* = \gamma d / U \approx 14$, $\alpha \approx 2.25$ and Lyapunov exponent $\lambda = 0.1$, obtained in Section~\ref{sec:def}, into equations~\eqref{eqsb1} and~\eqref{eqsb2}, we find $s^{(1)}_b < s^{(2)}_b$, since:
\begin{equation}
\lambda \ll {\gamma^*}(\gamma^*{s_0^*}^2 \textit{Pe})^{-1/(2\alpha + 1)} \ . \label{eqsepar}
\end{equation}
In particular we find $s^{(2)}_b \approx 5 s^{(1)}_b$.
This scale separation $s^{(1)}_b < s^{(2)}_b$ holds for moderate Péclet numbers, indicating that concentration elements first reach $s_b^{(1)}$ via shear-induced deformation and subsequently aggregate through diffusion to eventually attain the scale $s_b^{(2)}$ under exponential stretching at later times (provided that the shear induced mixing time $t_m$, here $t_m=0.33 d/U$, is smaller than the time for onset of exponential stretching $t_\epsilon =d/(U\lambda)$).

\begin{figure}
\centerline{\includegraphics[width=0.89\linewidth]{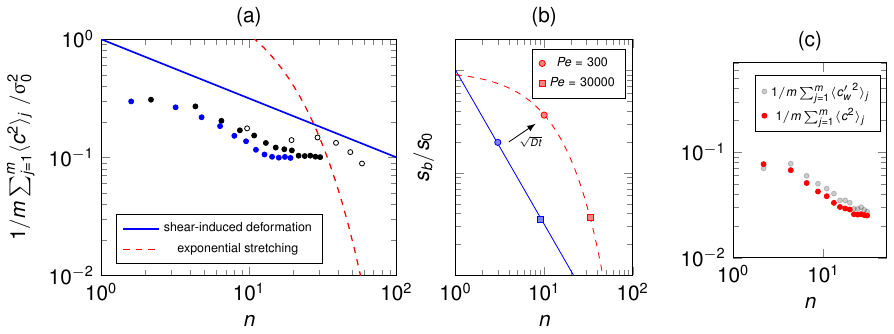}}
\caption{(a) Comparison between simulations (black marks: $R/d=4.7$; blue marks: $R/d=3.1$), the prediction for random transverse aggregation (equation~\eqref{eqoverlap}), and the exponential stretching model (dashed lines). An additional simulation with double reactor length is shown (empty square marks), yet exponential decay is barely observed up to $n \approx 50$. (b) Predicted compression of concentration sheets according to equations~\eqref{eqs1} (blue solid line) and exponential stretching (red dashed line), with markers indicating the thickness reached at the mixing times $s_b^{(1)}$ (blue) and $s_b^{(2)}$ (red) for $\textit{Pe} = 3 \times 10^2$ and $3 \times 10^4$, respectively. (c) Verification of volumetric variance and pore-scale mixing term $\sim \langle c_{w,\textit{max}}^2 \rangle$.}
\label{fig12}
\end{figure}

Before being subjected to exponential stretching, the concentration field reorganizes through diffusion. To investigate the expected asymptotic behavior, we performed an additional simulation by doubling the reactor length to $L/d = 84$, which allows the computation of the variance decay up to $n \approx 50$. The results are shown in figure~\ref{fig12}. A slight change in the variance decay is observed, hinting at the transition to exponential stretching. However, the limited computational domain prevents a definitive confirmation.
For compact, finite sized porous reactors, where reactions typically occur within $n \approx 10$–$50$, we conclude that shear-induced deformation remains the dominant mixing mechanism.

It is instructive to compare the observed algebraic decay of variance at early times with experimental observations in similar random packed beds~\citep{heyman2021scalar, heyman2020stretching}, where chaotic flows produce an exponential decay of variance. The Lyapunov exponent ($\lambda \approx 0.17$) and the onset time of chaotic mixing ($t_m^{(2)} \sim 6$) are qualitatively consistent with our findings. However, the early-time algebraic decay is not observed in these experiments. This discrepancy can be attributed to the different injection protocol. We also note that, due to the much higher Péclet numbers--which in those experiments are one to two orders of magnitude greater than in our simulations--the scale separation predicted by equation~\eqref{eqsepar} is greatly reduced, making any possible finite time shear-deformation less pronounced compared to exponential deformation.

To better highlight the $\Pen$-dependence of the scale separation, we can rewrite equation~\eqref{eqsepar} explicitly in terms of $\Pen$. Then to achieve a relevant difference between $s^{(1)}_b$ and $s^{(2)}_b$, e.g. $s^{(2)}_b > 4 s^{(1)}_b$ as in present simulations, the Péclet numbers must satisfies the condition:
 \begin{equation}
 \textit{Pe} < \frac{{\gamma^*}^{2\alpha}}{ {s_0^*}^2 (4^2 \lambda)^{2\alpha+1}} \ , \label{eqsepar2}
\end{equation}
which depends on the shear exponent $\alpha$. As an example, taking the characteristic pore size $s_0^*=1$ as initial concentration thickness, $\lambda=0.1$ and $\gamma^*\approx 10$, for strong non-linear shear ($\alpha=2$) we find $\Pen < 10^3$ while for linear shear ($\alpha=1$) we have $\Pen < 2 \ 10^1$.  Equation~\eqref{eqsepar2} can be interpreted as follows: shear dominates mixing at finite times if the Péclet number is substantially smaller than the ratio between the timescales characterizing the onsets of exponential mixing and shear-induced deformations.

\subsection{Mixing with surface reaction~\label{sec:rea}}

\begin{figure}
  \centerline{\includegraphics[width=0.99\linewidth]{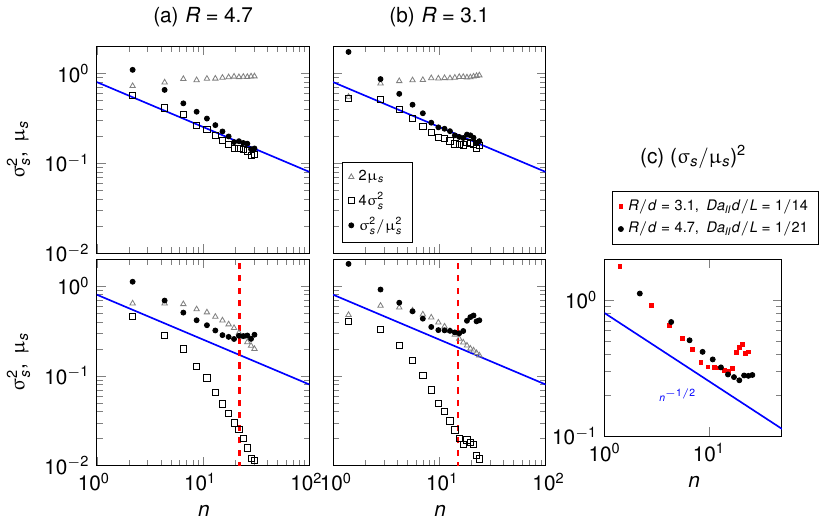}}
  \caption{Fluid-solid interfacial concentration mean $\mu_s$ and variance $\sigma_s^2$ evolution for (a) $R/d=4.7$ and (b) $R/d=3.1$ in an inert (top panels) and reactive (lower panels) media. The ratio $\sigma_s^2/\mu_s^2$ decays similarly, following equation~\eqref{eqoverlap} (solid blue lines) in the inert and reactive cases up to the characteristic reactive time $t_k=1/k$ (vertical red dashed lines in the bottom panels). After $t_k$ the variance-mean ratio at the reactive sites $\sigma_s^2/\mu_s^2$ saturates. The comparison of  $\sigma_s^2/\mu_s^2$ among the two two $L/d$ cases, reported in panel (c), shows that, provided the same $\textit{Da}_\textit{II}$ (see also table~\ref{tab1}) mixing is more effective for longer reactors, as predicted by equation~\eqref{eqreavar}.}
\label{fig13}
\end{figure}

In the presence of a reactive surface, such as the fluid–catalyst particle interface ($\textit{Da}_\textit{II} > 0$), the homogenization of surface concentration and local reaction is governed by the transverse overlap-driven mixing mechanism that determines the scaling of the volumetric variance in ~\eqref{eqoverlap}. Indeed, we observe that, for both inert and reactive media, the variance of concentration at the fluid–solid interface, $\sigma_s^2$, normalized by the mean $\mu_s$, follows:
\begin{equation}
\sigma_s^2/\mu_s^2 \ (t<t_k) = A n^{-1/2}/\mu_s^2(0) \propto ({\gamma^* \Pen})^{-\alpha/(2\alpha+1)} (t_D/t)^{1/2} \ , \label{eqrscale}
\end{equation}
up to the characteristic reaction time $t_k = 1/k_v$ (with $t_D=d^2/D$ a characteristic diffusion time). In the reactive case, we measure slightly higher values, but the temporal trend remains the same. 
At later times ($t > t_k$), $\sigma_s^2 / \mu_s^2$ appears to saturate for the reactive case (filled marks lower panels of Figure~\ref{fig13}). This effect is dictated by the local depletion of concentration at the reactive surface, which counterbalances mixing.

Then, by assuming that this mechanism suppresses mixing at the reactive surface for long times, we can make use of equations~\eqref{eqrscale} at time $t_k$ to estimate the variance-mean ratio at long times:
\begin{equation}
\sigma^2_s/\mu^2_s (t>t_k) \propto  ({\gamma^* \Pen})^{-\alpha/(2\alpha+1)} {\mbox{\textit{Da}}_\textit{I}}^{1/2} \ , \label{eqreavar}
\end{equation}
which again is consistent with the \Pen-scaling of second moments of return times in chaotic restart models~\citep{aquino2023fluid}, provided one sets $\alpha=1$. This consistency shows that both random aggregation models and chaotic restart models can in principle capture the heterogeneous distribution of local reactivities at the fluid-solid surface in reactive porous media.
Figure~\ref{fig13}, panel (c), illustrates this comparison for particle sizes $R/d = 3.1$ and $R/d = 4.7$, corresponding to $L/d = 28$ and $L/d = 42$, respectively (see Table~\ref{tab1}), where $\textit{Da}_\textit{I} = 3.33$ and $2.22$, respectively. We note that a lower $\textit{Da}_\textit{I}$ number (slower reaction) allows more mixing to take effect.

For design purposes, it is useful to highlight that  $\sigma^2_s \propto \mu^2_s {\mbox{\textit{Da}}_\textit{II}}^{1/2} (d/L)^{1/2}$ and that surface reaction homogenization (low $\sigma_s^2$) increases with the system length-to-particle-size ratio $L/d$. Thus,
for a given catalyst reaction rate $k_v$, optimal reactor design should (i) achieve near-complete conversion by maintaining $\textit{Da}_\textit{II} = k_v L / U = O(1)$, and (ii) promote effective mixing (low $\sigma_s$) by either reducing particle size $d$ or proportionally increasing both the flow rate $U$ and reactor length $L$. In practice, reducing particle size is often challenging due to production constraints and the risk of pore clogging, which limits reactor efficiency. Increasing flow rate and reactor length, on the other hand, offers a more practical strategy—a strategy that we have successfully validated in recent microreactor experiments~\citep{magson2024sust}.

\section{Conclusions~\label{sec:con}}

In this study, we investigate scalar transport and mixing in porous media at finite times by combining Lagrangian and Eulerian analyses based on a lattice-Boltzmann solver, to unveil the interplay between flow-induced deformation, diffusion, and reactive processes.  Although the mixing is chaotic, shear-induced deformation can dominate the early-time mixing dynamics, by compressing concentration regions transversely and promoting longitudinal aggregation, which enhances mixing. Transverse diffusion enables overlap between concentration sheets, leading to the formation of partially overlapping lamellar bundles and homogenization at the pore scale.
These mechanisms govern the evolution of both volumetric and surface concentration variances, producing a characteristic algebraic decay ($\sigma^2 \sim t^{-1/2}$) as verified by numerical simulations. These dynamics are observed when the Péclet number is much smaller than the ratio between the characteristic timescale of exponential mixing and that of shear-induced deformation, a condition that can be satisfied in small, compact fixed-bed reactors.

The aggregation dynamic, which combines correlated longitudinal with random transverse overlaps, explains the slower decay of volumetric variance compared to purely random three-dimensional mixing. This model highlights the critical role of shear-induced deformation in porous media at finite times. Although exponential stretching becomes theoretically relevant at later times, it plays only a secondary role here, because shear-driven mechanisms dominate over the timescales relevant for transport and reaction at moderate Péclet numbers.

In reactive systems, the homogenization of surface concentrations at fluid-catalyst interfaces mirrors the volumetric mixing trends, with the surface-concentration variance following the same scaling up to the characteristic reaction time. The homogenization of the surface concentration follows the \Pen-scaling $\sigma_s^2 \propto (\gamma^* \Pen)^{-\alpha/2(\alpha+1)}$, which enables the extension of shear-induced chaotic restart models~\citep{aquino2023fluid} beyond the linear shear regime ($\alpha = 1$). 

In porous systems where the divergence of chaotic trajectories is moderate (Lyapunov exponent $\lambda \approx 0.1$), the Péclet number is moderately high ($\Pen = O(10^2)$), and reaction is comparably slow to molecular diffusion ($\mbox{\textit{Da}}_\textit{I}= O(10^1)$),  strong shear ($\alpha \approx 2$) can dominate mixing at finite times and control the asymptotic pore-scale surface reaction.

\section*{Aknowledgements}
 
We thank the European Union's H2020 research and innovation program under grant agreement no. 951801 (MOST H2020-EIC-FETPROACT-2019-951801). The computations were enabled by resources provided by the National Academic Infrastructure for Supercomputing in Sweden (NAISS), partially funded by the Swedish Research Council through grant agreement no. 2022-06725.
The authors thank Daniel Lester for the fruitful discussion about the computation of the fluid deformation employed in the manuscript.

\section*{Declaration of Interests} 
The authors report no conflict of interest.

\section*{Appendix~\label{sec:app}}

The computation of the velocity gradient tensor $\boldsymbol{\epsilon}'(t)$ follows the procedure described in~\cite{lester2018fluid}. 
The moving Protean frame $\boldsymbol{x}'$ and the fixed Cartesian frame $\boldsymbol{x}$ are related via the rotation matrix $\mathsfbi{Q}(t)$ as:
\begin{equation}
\boldsymbol{x}'=\boldsymbol{x}_0(t)+\mathsfbi{Q}(t)^\mathrm{T} \boldsymbol{x} \ , \label{transf}
\end{equation}
with $\boldsymbol{x}_0(t)$ an arbirtary translation vector. The deformation tensor $\mathsfbi{F}(t)$ defines the deformations along a Lagrangian trajectory based on the velocity gradient tensor $\boldsymbol{\epsilon}(t) = \{ \boldsymbol{\nabla} \boldsymbol{u}[\boldsymbol{x}(t,\boldsymbol{X})] \}^\mathrm{T}$:
\begin{equation}
\frac{\mathrm{d}\mathsfbi{F}}{\mathrm{d} t}= \boldsymbol{\epsilon}(t) \mathsfbi{F}(t) \ . \label{eqdef0}
\end{equation}
The velocity gradient tensor computed along the Protean frame is:
\begin{equation}
\boldsymbol{\epsilon}'(t)=\mathsfbi{Q}(t)^\mathrm{T}\boldsymbol{\epsilon}(t)\mathsfbi{Q}(t) + \mathsfbi{A}(t) \ , \label{eqep}
\end{equation}
composed by two terms, the second of which quantifies the contribution of the moving reference frame:
\begin{equation}
\mathsfbi{A}(t)= \dfrac{\mathrm{d} \mathsfbi{Q}(t)^\mathrm{T} }{\mathrm{d} t} \ \mathsfbi{Q}(t) \ . \label{eqA}
\end{equation}
By applying the correct transformation~\eqref{transf} and exploiting the definitions~\eqref{eqep} and~\eqref{eqA}, it is possible to render the velocity gradient tensor upper-triangular and redefine equation~\eqref{eqdef0} along the Protean frame:
\begin{equation}
\frac{\mathrm{d}\mathsfbi{F}'}{\mathrm{d} t}= \boldsymbol{\epsilon}'(t) \mathsfbi{F}'(t) \ , \label{eqdef1}
\end{equation}
which is easily solvable for $\mathsfbi{F}'(t)$ with sequential integration, thus allowing the explicit definition of the shear and stretching components determining deformation. 

The orientation along the Protean frame is obtained via the rotation matrix $\mathsfbi{Q}(t)=\mathsfbi{Q}_1(t)\mathsfbi{Q}_2(t)$. Given the unit rotation axis $\boldsymbol{q}(t)$ and angle $\theta(t)$:
\begin{eqnarray}
\boldsymbol{q}(t) = \dfrac{1}{\sqrt{v_2^2+v_3^2}} (0, u_3,  -u_2)^\mathrm{T} \ , \\
\cos \theta(t) = \dfrac{u_1}{v}\ ,
\end{eqnarray}
where $\boldsymbol{v}(t,\boldsymbol{X})= (u_1,u_2,u_3)^\mathrm{T}$ is the instantaneous velocity  of magnitude $v$ along the computed Lagrangian trajectories, the first rotation matrix is defined as:
\begin{equation}
\mathsfbi{Q}_1(t) = \cos \theta(t) \mathsfbi{I} + \sin \theta(t) (\boldsymbol{q}(t)_\times)^\mathrm{T} + (1-\cos\theta(t)) \boldsymbol{q}(t) \otimes \boldsymbol{q}(t) \ . \label{eqQ1}
\end{equation}
In the above~\eqref{eqQ1}, $\mathsfbi{I}$ is the identity matrix, the pedix $\times$ denotes the cross-product matrix while $\otimes$ is the tensor product. The first rotation of the velocity gradient tensor produces the rotated tensor $\boldsymbol{\epsilon}^{(1)}(t)$:
\begin{equation}
\boldsymbol{\epsilon}^{(1)}(t) =\mathsfbi{Q}_1(t)^\mathrm{T} \boldsymbol{\epsilon}(t) \mathsfbi{Q}_1(t) \ . \label{eqe1}
 \end{equation}

The second rotation, applied via $\mathsfbi{Q}_2(t)$, is defined via the angle $\psi(t)$:
\begin{equation}
\mathsf{Q}_2(t) = \Bigg ( \ \ 
\begin{matrix}
0 & 0 & 0 \\
0 & \cos \psi(t)&  -\sin \psi(t) \\
0 & \sin \psi(t)&  \cos \psi(t) \\
\end{matrix}
\ \ \Bigg )  \ .
\end{equation}
The final tensor is provided by~\eqref{eqep}, which can be rewritten as: 
\begin{equation}
\boldsymbol{\epsilon}'(t) =\mathsfbi{Q}_2(t)^\mathrm{T} \boldsymbol{\epsilon}^{(1)}(t) \mathsfbi{Q}_2(t) \ . \label{eqe2}
\end{equation}
The upper-triangular condition on $\boldsymbol{\epsilon}'(t)$ is enforce by choosing $\psi(t)$ such that it
satisfies the following ordinary differential equation:
\begin{equation}
\dfrac{\mathrm{d}\psi(t)}{\mathrm{d} t} = a(t) \cos^2\psi(t) + b(t) \cos^2\psi(t) +c(t) \cos \psi(t) \sin \psi(t) \ , \label{eqode}
\end{equation}
where the $a(t)$, $b(t)$ and $c(t)$ coefficients are so defined:
\begin{eqnarray}
a(t) = \epsilon^{(1)}_{32} + \dfrac{u_2}{v+u_1}\epsilon^{(1)}_{31} -  \dfrac{u_3}{v+u_1}\epsilon^{(1)}_{21}  \ ,\\
b(t) = \epsilon^{(1)}_{23} + \dfrac{u_2}{v+u_1}\epsilon^{(1)}_{31} -  \dfrac{u_3}{v+u_1}\epsilon^{(1)}_{21}  \ ,\\
c(t) = \epsilon^{(1)}_{22} + \epsilon^{(1)}_{33}  \ ,\\
\end{eqnarray}
with $\epsilon^{(1)}_{ij}$ the $i,j$ component of the tensor $\boldsymbol{\epsilon}^{(1)}$.
\cite{lester2018fluid} showed that the differential equation~\eqref{eqode}, whilst allowing multiple solutions, admits an attracting trajectory for a specific value of the initial angle condition $\psi(0)$. Such a value can be estimated at short times, via the minimization of $\mathrm{d} \dot{\psi} / \mathrm{d} \psi$, where $\dot{\psi}=\mathrm{d}\psi/\mathrm{d} t$. Once $\psi_0$ is identified, the solution of~\eqref{eqode} provides the temporal solution of $\psi(t)$, $\mathsfbi{Q}_2(t)$ along a lagrangian trajectory. From equation~\eqref{eqe2}, the deformation tensor $\mathsfbi{F}'(t)$ can be calculated sequentially as:
\begin{eqnarray}
F'_{11} = \dfrac{v(t)}{v(0)} \\
F'_{22} = \exp \bigg [   \int_0^t \epsilon'_{22}(t)\ \mathrm{d} t \bigg ] \\
F'_{33} = \exp \bigg [   \int_0^t \epsilon'_{33}(t)\ \mathrm{d} t \bigg ] \\
F'_{12} = v(t) \int_0^t \dfrac{\epsilon'_{12}(t) F'_{22}(t)}{v(t)}\ \mathrm{d} t \\
F'_{23} = F'_{22}(t) \int_0^t \dfrac{\epsilon'_{23}(t) F'_{33}(t)}{F'_{22}(t)}\ \mathrm{d} t \\
F'_{13} = v(t) \int_0^t \dfrac{\epsilon'_{12}(t) F'_{23}(t)+\epsilon'_{13}(t) F'_{33}(t)}{v(t)}\ \mathrm{d} t \\
\end{eqnarray}

 \begin{figure}
  \centerline{\includegraphics[width=0.99\linewidth]{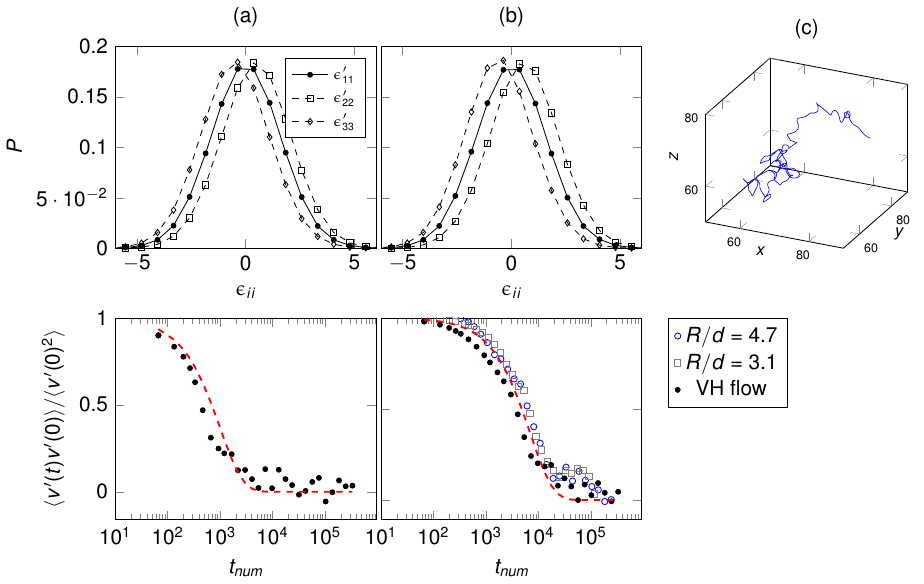}}
  \caption{Upper panels: probability distribution functions $P$ of the diagonal components of the velocity gradient tensor $\epsilon'_{ii}$ for a characteristic decorrelation time $t'_\textit{num}=10^3$ (a) and  $t'_\textit{num}=10^4$ for the VH flow. The mean value $\langle \epsilon'_{22}\rangle$ identifies the Lyapunov exponent $\lambda=0.48$ (a) and $\lambda=0.57$ (b). Lower panels: velocity autocorrelation function $\langle v'(t)v'(0)\rangle/\langle v'^2(0)\rangle$ (with $v'(t) = v(t)-\langle v(t) \rangle$ the velocity magnitude fluctuation) identifying the decorrelation characteristic time $t'_\textit{num}=10^3$ (a) and $t'_\textit{num}=10^4$ (b). The red dashed lines represent the exponential decay $\langle v'(t)v'(0)\rangle/\langle v'^2(0)\rangle = \exp(-t_\textit{num}/t'_\textit{num})$. In the bottom right panel the autocorrelation functions for the reactor simulations $R/d=4.7$ and $R/d=3.1$. Right panel (c): a trajectory computed on the VH flow.}
\label{figA}
\end{figure}

We test the procedure described above in the variable helicity (VH) flow as described in~\cite{lester2018fluid}. The results are shown in figure~\ref{figA} and a typical trajectory, computed via a Lagrangian tracking algorithm, is presented in the right panel (c). We are particularly interested in the effect of the temporal resolution of trajectory solver on the results accuracy. Once the temporal resolution is chosen, the VH flow supports trajectories that decorrelate over a certain characteristic time $t'_\textit{num}$, see lower panels in figure~\ref{figA}. We find that to the temporal resolutions  $t'_\textit{num} = 10^3$ and $t'_\textit{num} = 10^4$ correspond the mean diagonal components $\langle \epsilon'_{ii} \rangle = (0, 0.48,-0.48)$ and $\langle \epsilon'_{ii} \rangle = (0, 0.57,-0.57)$, respectively. The latter value is not so far from the theoretical maximum of the Lyapunov exponent $\ln(2)$, indicating a good temporal resolution for $t'_\textit{num}= 10^4$ . The same resolution is used for the simulations of the trajectories in the reactor, $R/d=4.7$ and $R/d=3.1$, as shown in the lower panel (b) of figure~\ref{figA}.

\bibliographystyle{jfm}
\bibliography{jfm-v5}

\end{document}